\documentclass[12pt,preprint]{aastex}

\newcommand{\ASCA}{{\sl ASCA}}
\newcommand{\Chandra}{{\sl Chandra}}

\newcommand{\mmsnr}{MM~SNR}
\newcommand{\mmsnrs}{MM~SNRs}
\newcommand{\ic}{IC~443}
\newcommand{\kes}{Kes~27}
\newcommand{\kTe}{$kT_{\mathrm{e}}$}
\newcommand{\kTz}{$kT_z$}
\newcommand{\Hea}{He$\alpha$}
\newcommand{\Heb}{He$\beta$}
\newcommand{\Her}{He$\gamma$}
\newcommand{\Lya}{Ly$\alpha$}
\newcommand{\Lyb}{Ly$\beta$}
\newcommand{\ratio}{Ly$\alpha$/He$\alpha$}
\newcommand{\ntunit}{$\mathrm{s~cm^{-3}}$}
\newcommand{\Msun}{M\mbox{$_\odot$}}
\newcommand{\Nh}{$N_{\mathrm{H}}$}

\slugcomment{This is the astro-ph version}

\shorttitle{ASCA Observations of Mixed-morphology SNRs}
\shortauthors{Kawasaki et al.}

\begin{document}


\title{
Ionization States and Plasma Structures of Mixed-morphology SNRs
Observed with \ASCA}


\author{Masahiro Kawasaki\altaffilmark{1}\altaffilmark{2},
	Masanobu Ozaki\altaffilmark{1}, Fumiaki Nagase\altaffilmark{1},
	Hajime Inoue\altaffilmark{1},
	and Robert Petre\altaffilmark{3}}

\altaffiltext{1}{The Institute of Space and Astronautical Science,
Sagamihara, Kanagawa, 229-8510, Japan}
\altaffiltext{2}{Mitsubishi Heavy Industries, LTD.,
Nagoya Guidance and Propulsion Systems Works,
1200, Higashi-tanaka, Komaki, Aichi, 485-8561, Japan}
\altaffiltext{3}{Laboratory of High Energy Astrophysics, Code 662,
NASA/Goddard Space Flight Center, Greenbelt, MD 20771;
rob@lheapop.gsfc.nasa.gov}



\begin{abstract}
We present the results of a systematic study using \ASCA\ of the
ionization state for six ``mixed-morphology'' supernova remnants
(\mmsnrs): \ic, W49B, W28, W44, 3C391, and Kes~27.  \mmsnrs\ show
centrally filled thermal X-ray emission, which contrasts to
shell-like radio morphology, a set of characteristics at odds with the
standard model of SNR evolution (e.g., the Sedov model).  We have
therefore studied the evolution of the \mmsnrs\ from the ionization
conditions inferred from the X-ray spectra, independent of X-ray
morphology.  We find highly ionized plasmas approaching ionization
equilibrium in all the \mmsnrs.  The degree of ionization is
systematically higher than the plasma usually seen in shell-like SNRs.
Radial temperature gradients are also observed in five remnants, with
cooler plasma toward the limb. In \ic\ and W49B, we find a plasma
structure consistent with shell-like SNRs, suggesting that at least
some \mmsnrs\ have experienced similar evolution to shell-like SNRs.
In addition to the results above, we have discovered an
``overionized'' ionization state in W49B, in addition to that
previously found in \ic. Thermal conduction can cause the hot interior
plasma to become overionized by reducing the temperature and density
gradients, leading to an interior density increase and temperature
decrease.  Therefore, we suggest that the ``center-filled'' X-ray
morphology develops as the result of thermal conduction, and should
arise in all SNRs.  This is consistent with the results that \mmsnrs\
are near collisional ionization equilibrium since the conduction
timescale is roughly similar to the ionization timescale.  Hence, we
conclude that \mmsnrs\ are those that have evolved over
$\sim$10$^4$~yr.  We call this phase as the ``conduction phase.''
\end{abstract}




\section{Introduction}
\label{intro}

The most prominent characteristics of the plasma in a supernova
remnant (SNR) are its high electron temperature ($\sim$10$^7$~K) and
low density ($\le$1~cm$^{-3}$). It takes a substantial fraction of the
life of a SNR ($10^{4-5}$~yr) for ions in such an optically-thin
thermal plasma to reach collisional ionization equilibrium (CIE) with
the electrons heated by the shock wave. Such an ``underionized'' state
is often referred to as non-equilibrium ionization (NEI) and has been
observed in many SNRs.  Thus, investigating the range of ionization
states in a SNR provides significant information about its evolution.
The objects of the present study are ``mixed-morphology'' SNRs
(\mmsnrs) whose formation mechanism is not yet well understood. The
properties of \mmsnrs\ have been most succinctly described in
\citet{rho98}. They have a shell-like radio morphology, and their
steep radio spectrum is characteristic of synchrotron radiation from
electrons accelerated by the shock wave.  In contrast, the X-ray
morphology of \mmsnrs\ is centrally peaked, with little or no limb
brightening.  Unlike the centrally peaked synchrotron radiation from
pulsar wind nebulae such as the Crab Nebula, the X-ray emission is
thermal, showing line emission from abundant metals such as Mg, Si, S,
and Fe.  In general, thermal X-rays are thought to arise mainly from
the shock-heated and compressed material behind the forward blast wave
that would give rise to a shell-like morphology. Thus, the properties
of \mmsnrs\ are inconsistent with the initial $\sim$10$^4$~yr of the
standard model of SNR evolution (i.e., the Sedov model), when the
shock front is moving fast enough to heat ambient material to X-ray
emitting temperatures. However, \mmsnrs\ represent about 20\% of the
X-ray detected Galactic SNRs, suggesting that the conditions for
creating such remnants are not rare.
A large number of them are interacting with molecular clouds and/or
\ion{H}{1} clouds. For example, OH maser emission at 1720~MHz is
detected from seven \mmsnrs, a remarkably high fraction
\citep{zadeh03}.

Two scenarios have been invoked to explain the X-ray
morphology of  \mmsnrs. 
One is that the interior X-ray emission arises from the
gas evaporated from shocked clouds \citep{white91}, implying that
\mmsnr\ have evolved under special conditions. The other is that 
as a SNR evolves, the temperature and density of the 
hot interior plasma gradually becomes uniform through thermal conduction,
and once the remnant has entered the ``shell forming'' stage in which the 
forward shock no longer heats ambient material to X-ray emitting temperatures,
it becomes visible as centrally brightened X-ray emission
\citep{cox99,shelton99b}.  However, the formation
mechanisms of \mmsnrs\ are still controversial.  We investigate the
ionization degree of the plasma to study the evolution of \mmsnr\
since the ionization state of a remnant can yield an age estimate that
is independent of the X-ray morphology.

In this paper, we first describe our selection of targets for
analysis of ionization state and plasma structure
(section~\ref{selection}).  Section~\ref{analysis} presents the
analysis and results of individual remnants, the details of which are in 
\citet{kawasakiphd}. In section~\ref{discuss},
we summarize the results of the \mmsnrs\ through comparison with
shell-like SNRs.  We then discuss possible mechanisms for the creation
of the newly-found ``overionized'' plasma and \mmsnrs\ in general.  We
conclude that both the plasma structure and ionization conditions can
be explained by thermal conduction.  We finally present the
implications of our findings for the overall picture of SNR evolution.

\section{Target Selection}
\label{selection}

In order to investigate the ionization conditions in \mmsnrs, we adopted
the following method to supplement global NEI model fits.  First, we estimate
the electron temperature, \kTe, from the continuum shape of the X-ray
spectrum.  Next, we measure the intensity ratio of the H-like K$\alpha$
(hereafter \Lya) line to the He-like K$\alpha$ (hereafter \Hea) line
for the heavy elements, which provides a measure of the degree of ionization
as the ionization temperature, \kTz. Comparison between \kTe\ and
\kTz\ directly gives the ionization state of the SNR from its X-ray
spectrum. This method is to produce robust results, then 
we need to utilize data from
an X-ray spectrometer that resolves the \Lya\ line from the \Hea\ line. 
One such instrument is the Solid-state Imaging Spectrometer (SIS) onboard
\ASCA, which has energy resolution of 5\% (FWHM) at 1.5~keV.
However, the SIS energy resolution degraded over the
course of the mission as a consequence of
accumulated radiation damage \citep{yamashita99}.  Therefore, only data
from early \ASCA\ observations (from the launch in 1993 through 1994) are
suitable for this analysis.

Table~\ref{list} tabulates the \ASCA\ observations of candidate \mmsnrs\
from \citet{rho98}.  Among the candidates, MSH~11-54 (G292.2+1.8),
their group ``F'' remnant,
is excluded because it is considered to be
a reverse-shocked, ejecta-dominated young remnant, showing strong O
and Ne lines with a large velocity of 2,000~km~s$^{-1}$
\citep{goss79,murdin79}, and a faint shell in X-rays
\citep{park02}.
For the systematic study of ionization states, we used only
the \ASCA\ data. Thus, our criteria of selecting targets in this paper
are: (1) \mmsnr\ candidates assigned as A to E in Table~\ref{list},
and (2) SNRs observed with \ASCA\ from 1993 to 1994.  There are eight
SNRs satisfying these criteria.  However, the data from 3C400.2 and
MSH~11-61A, observed in 1994, are not suitable for our analysis as
mentioned in the notes of table~\ref{list}.  Therefore, we analyzed
the following six remnants: \ic\ (G189.1+3.0), W49B (G43.3$-$0.2), W28
(G6.4$-$0.1), W44 (G34.7$-$0.4), 3C391 (G31.9+0.0), and Kes~27
(G327.4+0.4).

\section{Data Analysis and Results}
\label{analysis}

For the series of analysis, we extracted all the data from \ASCA\
public archive, and screened them using the the NASA/GSFC
revision~2 standard data processing criteria.  We applied no Residual
Dark Distribution (RDD) correction because the early-mission
SIS data show few or no
radiation damage effects.  As most of these remnants are
located near the Galactic plane where the Galactic Ridge X-ray emission
(GRXE) is not negligible, we used source-free regions of the
individual observations for background subtraction.
Exceptions to this background subtraction approach were \ic\ and W49B,
for which we used blank-sky data: \ic\ is located in the
Galactic anti-center region, and the results for W49B are not
sensitive to the
background used \citep{hwang00}.

When investigating the ionization state, we used a model consisting of
a CIE plasma ({\tt VMEKAL} in XSPEC,
\citet{mewe85,kaastra92,liedahl95}) and several Gaussian components. A
CIE plasma model with metal abundances set to zero represents the
continuum shape (i.e., \kTe), while Gaussian components are used to
measure the line intensity ratios (\ratio) of heavy elements (i.e.,
\kTz).  We fixed the Gaussian widths at 10~eV for \Hea\ of Mg, Si, S,
and Ar, and 20~eV for \Hea\ of Ca, in order to account for the fact
that the \Hea\ lines are triplets.  The widths of all other lines were
fixed at 0.1~eV, corresponding functionally to monochromatic lines
compared with the SIS energy resolution.

For remnants whose angular size is larger than the SIS field of
view (FOV), we extracted spectra from a few regions and
investigated the plasma structure and ionization state of each.  Below, we
present the results for individual remnants.

\subsection{\ic}

The details of the analysis and results for \ic\ are described in
\citet{kawasaki02}; we briefly summarize the results here.  \ic\ has a
$\sim$45\arcmin\ diameter, and we extracted the spectra from two
regions: a central region where the X-ray surface brightness is
highest, and a region near the northern rim.  Both spectra require two
plasma components (1.0~keV and 0.2~keV); the emission of the 0.2~keV
plasma is stronger in the region near the shell than the center.  In
addition, the mean electron densities of the 1.0~keV and 0.2~keV
plasmas estimated from the emission measures are 1~cm$^{-3}$ and
4~cm$^{-3}$ respectively.  These results can be accounted for by a
simple model in which \ic\ has a hot (1.0~keV) interior surrounded by
a cool (0.2~keV) and denser shell.  In the high-temperature plasma,
the ionization temperature from the \ratio\ ratio of S is
$\simeq$1.5~keV, significantly higher than the electron temperature of
1.0~keV. Neither an additional, hotter plasma component nor a
multi-temperature plasma successfully accounts for this feature, and
we conclude that the interior hot plasma of \ic\ is overionized.

We attempted to verify this result using XMM-Newton data.  Unfortunately,
the observation of \ic\ suffers from high background; the available data
do not provide results as statistically accurate as those from \ASCA.

\subsection{W49B}

\begin{figure}
\epsscale{0.325}
\plotone{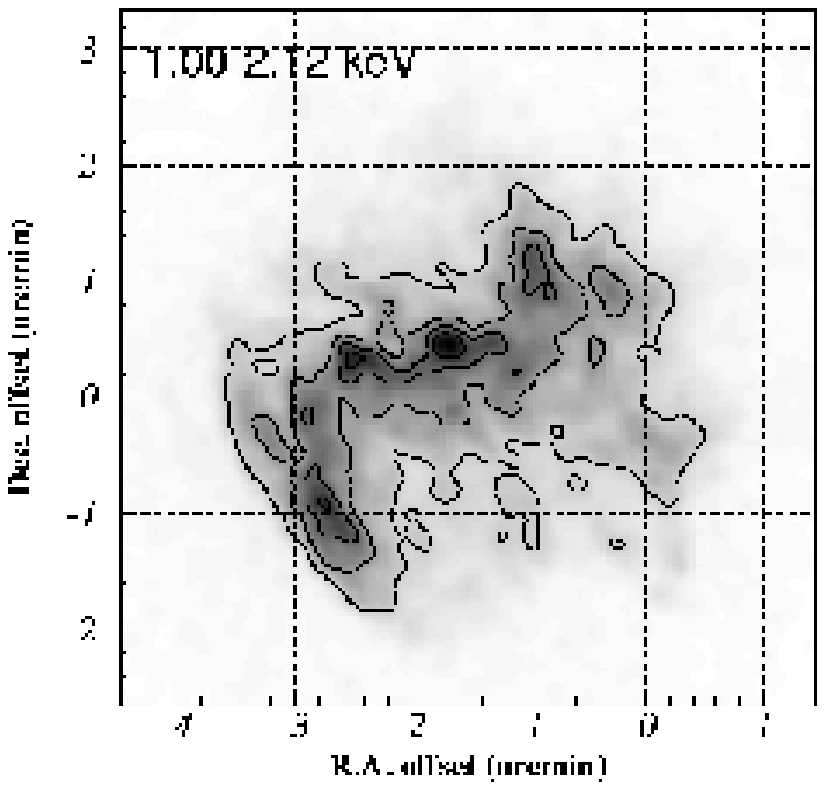}
\plotone{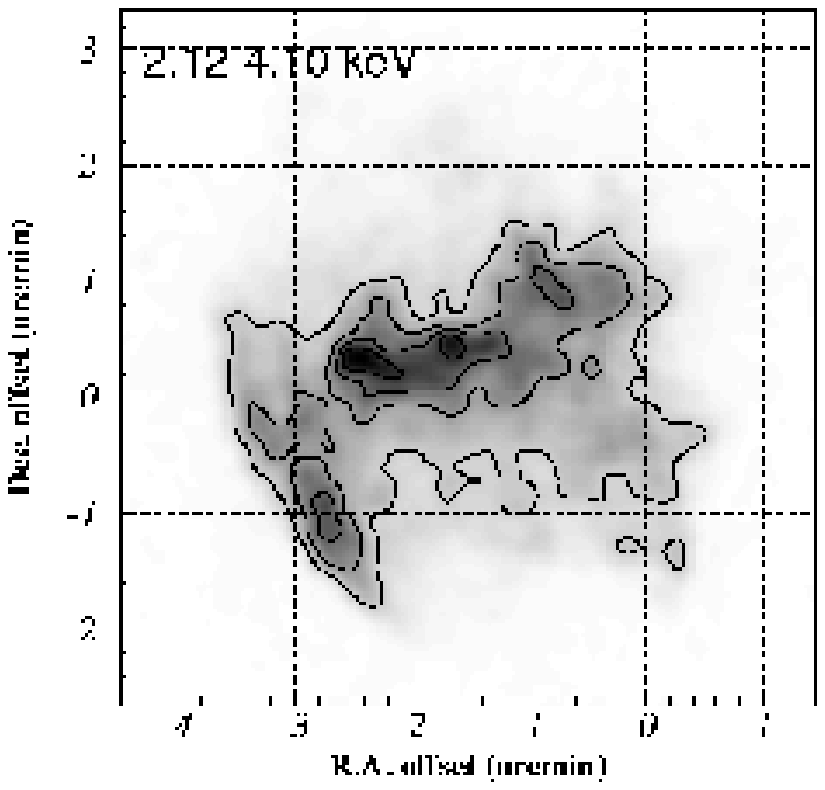}
\plotone{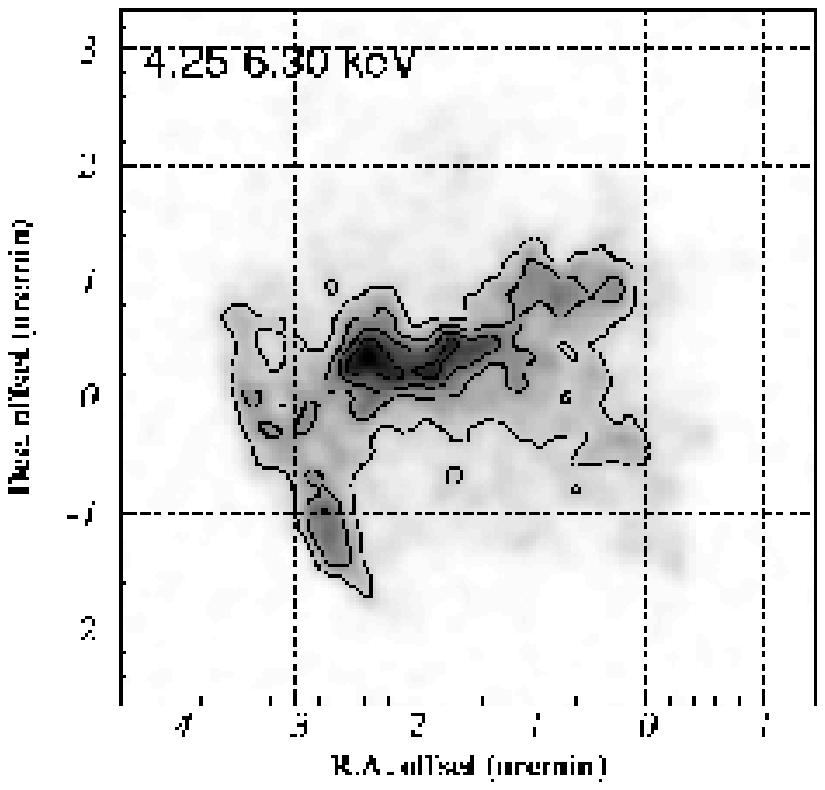}
\caption{\Chandra\ ACIS images of W49B in ({\it left}) 1.00--2.12~keV,
({\it middle}) 2.12--4.10~keV, and ({\it right}) 4.25--6.30~keV bands.
Grayscales are linear and the contour levels in each figure are 20\%,
40\%, 60\%, and 80\% of the peak surface brightness. Offset center in
each figure is R.A.=19\fh11\fm1\fs35,
Dec.=9\arcdeg06\arcmin7.64\arcsec (J2000). All images are derived from
\Chandra\ Supernova Remnant Catalog Web page
(http://hea-www.harvard.edu/ChandraSNR/).}
\label{w49bimg}
\end{figure}

\begin{figure}
\epsscale{0.5}
\plotone{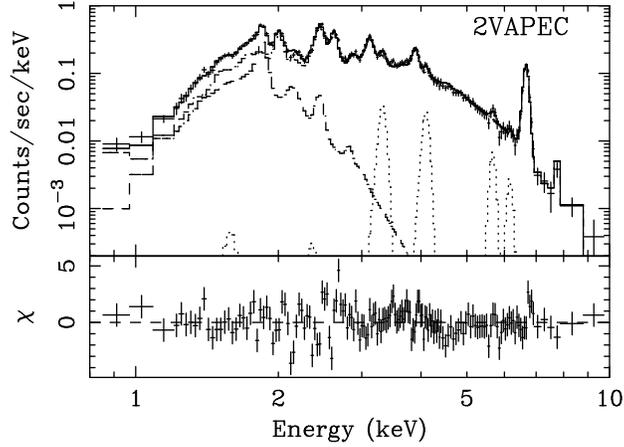}
\caption{\ASCA\ SIS spectra of the entire region of W49B with
the best-fit 2 CIE plasma and Gaussians model. 
Dashed, dash-dotted, and dotted lines represent the low-T plasma, 
high-T plasma, and additional Gaussian components
respectively. The lower panels show the residuals of the fits.}
\label{w49b2pla}
\end{figure}

W49B appears centrally brightened in X-rays and shell-like at radio
wavelengths, like other \mmsnrs. However, this remnant is thought to
be ejecta-enriched, and thus relatively young,
based on the high (3--5 times as solar) metal abundances
\citep{fujimoto95, hwang00}.
Since W49B is compact, with a size of only 4\arcmin$\times$3\arcmin\, we
used the \Chandra\ images with 0.5\arcsec\ spatial resolution to
ascertain the energy dependence of its morphology.  
Figure~\ref{w49bimg} shows the
\Chandra\ ACIS images in the 1.0--2.12~keV, 2.12--4.10~keV, and
4.25--6.30~keV energy bands.  They show that the highest energy image
is more centrally concentrated than those in the lower energy bands.
The SIS spectrum of the entire remnant in the 0.8--10.0~keV energy
band (2CCD mode data) requires at least a two-temperature
(0.24$^{+0.04}_{-0.02}$~keV and 1.70$^{+0.02}_{-0.04}$~keV) CIE plasma
({\tt VAPEC} in XSPEC, \citet{smith01}) with a large absorption column
density of 5.23$^{+0.11}_{-0.10}\times10^{22}$~cm$^{-3}$. Furthermore,
we had to add four Gaussian components to reproduce unusually strong
emission lines of Ar \Lya\ and Ca \Lya, whose center energies were
fixed at theoretical values of 3.324~keV and 4.105~keV respectively, as
well as Cr \Hea\ and Mn \Hea\ lines \citep{hwang00}.  The reduction of
$\chi^2$ by adding each \Lya\ line is more than 10 for 147 d.o.f, and
an F-statistic test gives a probability greater than 97\% indicating
that these lines are real.  The resultant $\chi^2$/d.o.f. is 233/146
(see Tab.~\ref{w49bparam} and Fig.~\ref{w49b2pla}).

\begin{figure}
\epsscale{1.0}
\plottwo{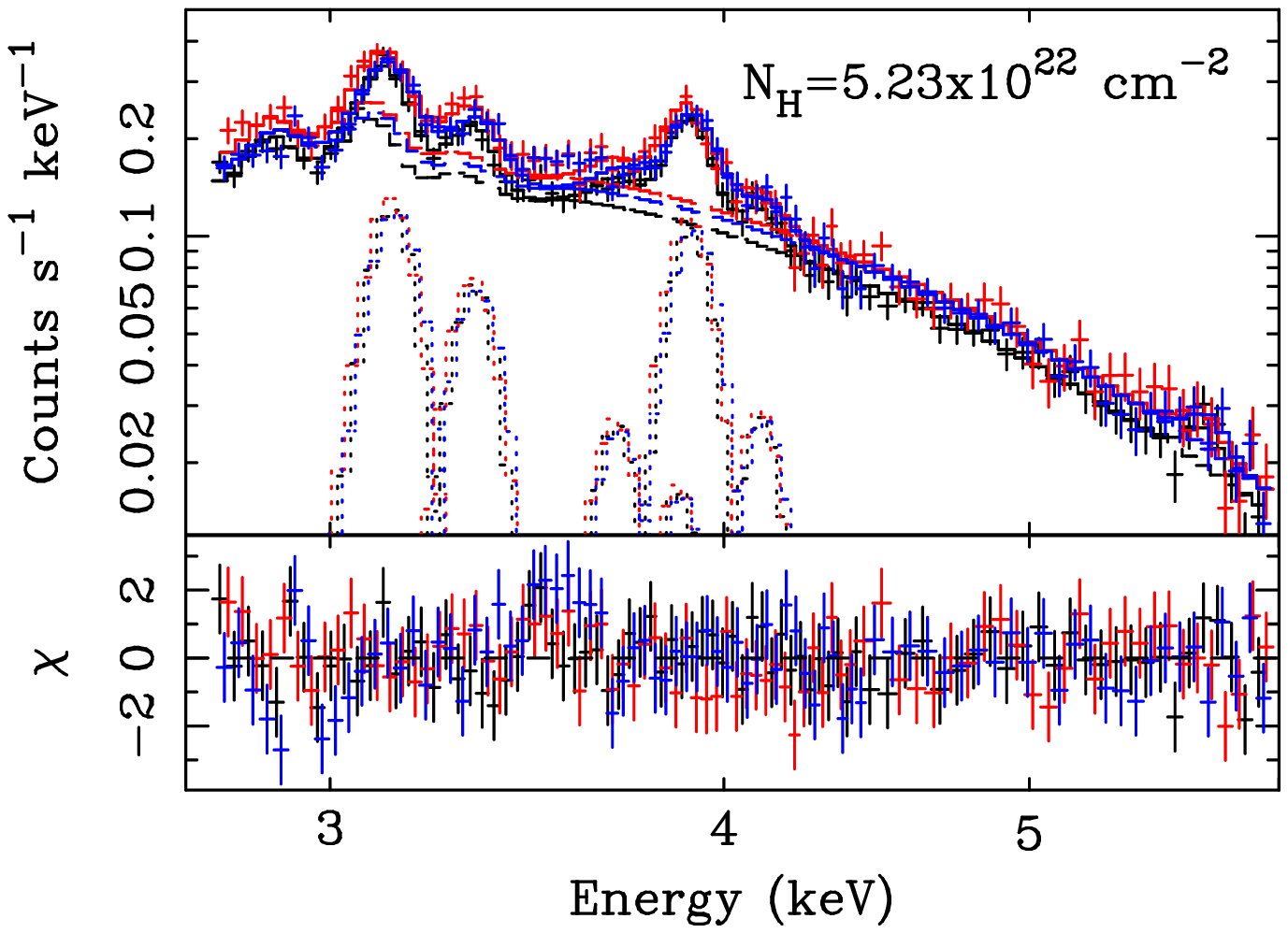}{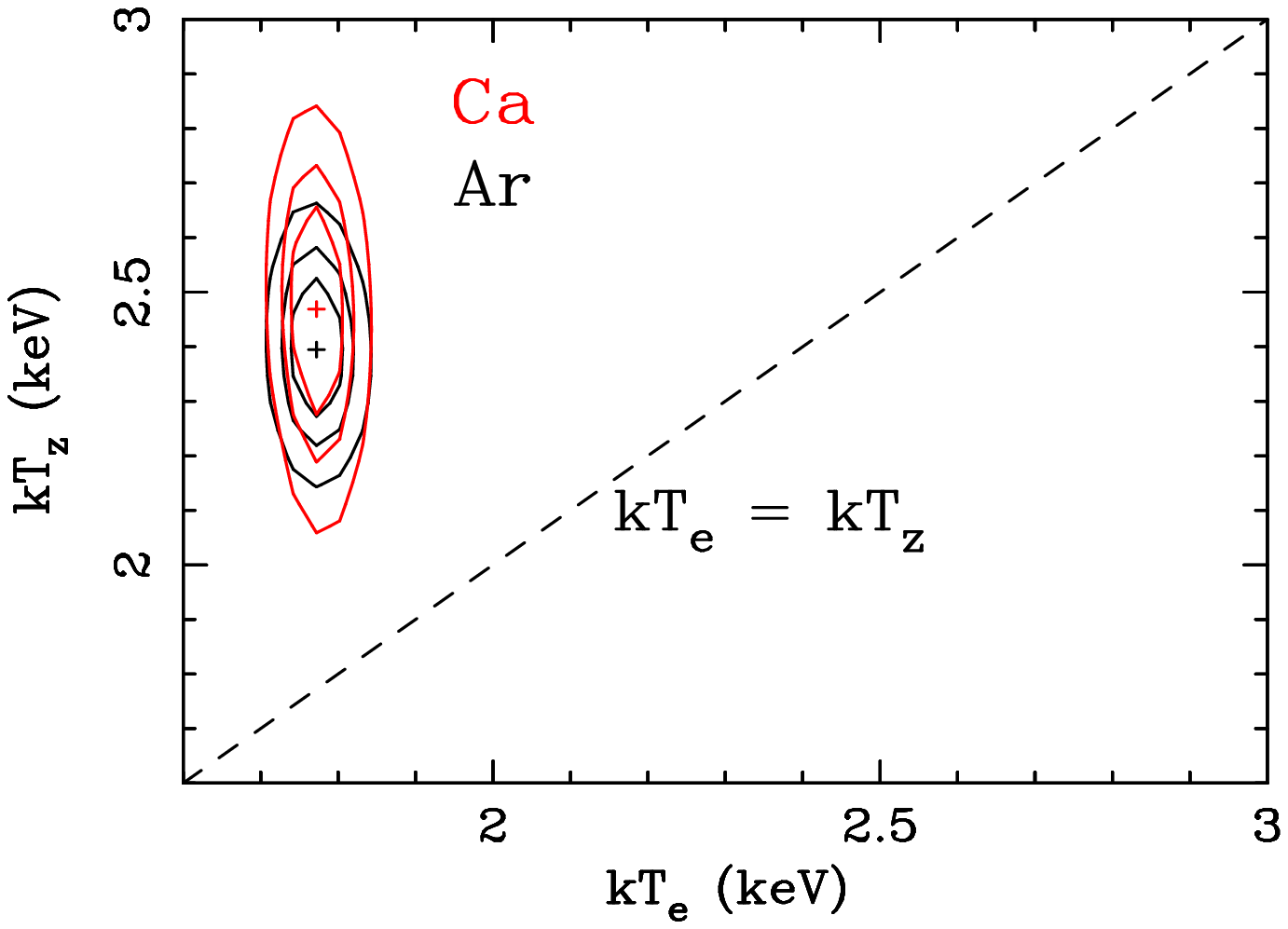}
\caption{{\it left}: \ASCA\ SIS spectra of W49B in 2.75--6.0~keV
energy band with the best-fit model of {\tt VMEKAL} (dashed) and 9
Gaussian lines (dotted).  The data in 2CCD mode are plotted in black,
and those in 1CCD mode are in red and blue.  The column density is
fixed to \Nh=5.23$\times$10$^{22}$~cm$^{-2}$.  The minimum value of
$\chi^2$/d.o.f. is 210/235. The lower panel shows the fit residuals.
There are residuals at $\sim$3.5~keV, which can be identified as a
He-like K$\alpha$ line of potassium.  {\it right}: Confidence contours
of ratio of the ionization temperature (\kTz) to the electron
temperature (\kTe) for Ar (black) and Ca (red) with a fixed column
density of 5.23$\times$10$^{22}$~cm$^{-2}$.  The confidence levels are
99\%, 90\% and 67\%.  The dashed line represents CIE.}
\label{w49bspec}
\end{figure}

We therefore evaluated the ionization temperatures of Ar and Ca and
compared them to the continuum temperature. In order to achieve a
better statistical accuracy, we added the two data sets taken in 1CCD
mode (13ksec and 19ksec SIS data) and fitted them simultaneously.  We
used the spectra in the 2.75--6.0~keV energy band since the
contribution of the low-temperature plasma is negligible at these
energies.  We applied a model consisting of the CIE plasma ({\tt
VMEKAL}) and 9 Gaussian components (\Hea, \Lya, \Heb, \Her\ of Ar,
\Hea, \Lya, \Heb, \Her\ of Ca, and \Hea\ of Cr) as shown in
Figure~\ref{w49bspec}~({\it left}). Line center energies of Gaussian
components were fixed at the theoretical values except for Ar \Hea, Ar
\Lya, Ca \Hea, and Ca \Lya.  For the W49B plasma with strong emission
lines, line blending can be a significant issue when measuring line
fluxes with a moderate resolution spectrometer like the SIS.  A
non-negligible flux from S H-like K$\beta$ (\Lyb) at 3.107~keV can
blend with Ar \Hea\ ($\sim$3.13~keV), as can
\Heb\ and \Her\ emission of S at 2.88~keV and 3.02~keV
respectively. In these fits, we fixed the S abundance to 3.9 solar as
derived from the 0.8--10.0~keV spectrum fitting, instead
of adding further Gaussian components. We ignored the Ar \Lyb\
(3.936~keV) flux, which could mildly contaminate (5--10\%) the Ca
\Hea.  For \Heb\ and \Her\ line blending, we modeled the \Her\ line
with an intensity relative to the \Heb\ line fixed at 0.6
\citep{hwang00}.  Both Ar and Ca, the \ratio\ ratio requires an 
ionization temperature
of $\sim$2.5~keV; this is higher than the $\sim$1.8~keV
electron temperature (see Fig.~\ref{w49bspec}~({\it right}) for the
confidence contours).  Adding a third plasma component with higher
($>1.8$~keV) temperature can in principle account for the strong
H-like lines of Ar and Ca, but it is required to have an implausible
combination of small emission measure and extremely large
($\gtrsim100$) abundances.  Therefore, we suggest that the
high-temperature plasma of W49B is ``overionized'', similar to
\ic. W49B is thus the second SNR showing overionization.

Our results may be compared with the detailed spectral analysis by
\citet{hwang00}. They showed the constraints on \kTz\ for the \ratio\
of Ar and Ca to be 2.2--2.7~keV. The values are higher than the \kTe\
of 1.7~keV inferred from the bremsstrahlung continuum.  Our results are
consistent with theirs, only our interpretation differs.

Taken together, the \Chandra\ images and \ASCA\ spectra indicate the
presence of central hot and outer cool plasmas. We thus suggest that
W49B consists of two typical plasma components (a 1.7~keV hot 
interior and a 0.24~keV cool shell), like that of the Sedov model, though
\citet{fujimoto95} suggested a possible stratification of ejecta.

The electron densities can be estimated from the emission measures to
be
\begin{eqnarray}
n_1 &=& 2.5\; \left( \frac{f_1}{0.64} \right)^{-1/2} \left(
         \frac{\theta}{2\arcmin} \right)^{-3/2} \left(
         \frac{d}{11.4~\mathrm{kpc}} \right)^{-1/2}
         ~\mathrm{cm^{-3}},\\
n_2 &=& 18\; \left( \frac{f_2}{0.36}
         \right)^{-1/2} \left( \frac{\theta}{2\arcmin} \right)^{-3/2}
         \left( \frac{d}{11.4~\mathrm{kpc}}
         \right)^{-1/2}~\mathrm{cm^{-3}},
\end{eqnarray}
for the high- and low-temperature components respectively, assuming
that the two components are in pressure equilibrium and the X-ray
emitting volume extends to the radio shell (which we approximate as a
spherical region with angular radius $\theta$=2\arcmin\ ).  Here,
$f_1$ and $f_2$ ($f_1 + f_2 = 1$) are the filling factors of these
components and $d$ is the distance to W49B.
This plasma structure, temperature and density distributions, is
consistent with the idealization of the Sedov model.

The X-ray emitting masses of the components can also be estimated:
\begin{eqnarray}
M_1 &=& 49 \left( \frac{f_1}{0.64} \right)^{1/2}
         \left( \frac{\theta}{2\arcmin} \right)^{3/2}
         \left( \frac{d}{11.4~\mathrm{kpc}} \right)^{5/2}~\Msun,
\label{w49b_m1}\\
M_2 &=& 2.0 \times10^2 \left( \frac{f_2}{0.36} \right)^{1/2}
         \left( \frac{\theta}{2\arcmin} \right)^{3/2}
         \left( \frac{d}{11.4~\mathrm{kpc}} \right)^{5/2}~
         \Msun
\label{w49b_m2}.
\end{eqnarray}
The masses of the high- and low-temperature plasmas are much higher
than that of any plausible progenitor, even if $d = 8$~kpc which is
favored by \citet{moffett94b}.  We should note here that the masses
are estimated assuming uniform density distributions within the
remnant.
The \Chandra\ images show some clumpy structures as well
as the extended components.  In fact, the mass of the high-temperature
plasma could be $\sim\Msun$ if the X-rays come only from
clumps \citep{hwang00}.  However, the extent of the 1.00--2.12~keV 
\Chandra\ image (see
Fig.~\ref{w49bimg}) indicates that the mass of the
low-temperature component is much higher than the ejecta mass.  
Coupled with the surprisingly close
proximity of the plasma to ionization equilibrium (in 
stark contrast to ejecta-dominated remnants (e.g.,G292.0$+$1.8 --
\citep{hughes94}), the high mass leads us to suggest that W49B
is not a young, ejecta-dominated, but an evolved \mmsnr.

The mass and ionization state of W49B are reasonably explained if the
plasma is dominated by the shocked ambient medium, but its large
abundances and high temperature (up to $\simeq$1.7~keV) strongly
suggests the the presence of ejecta.  Overabundances might be
explained if W49B exploded in the massive ($\sim$10$^{2}$~\Msun)
circumstellar cloud with large amount of heavy elements seeded by
previous SNRs.  The X-ray measured absorption column density is
5.23$^{+0.11}_{-0.10}\times$10$^{22}$~cm$^{-3}$, significantly higher
than the Galactic column density of 4.4$\times$10$^{22}$~cm$^{-3}$
measured in this direction using the 21cm and CO observations
\citep{dickey90,stark85}. The X-ray absorption is mainly due to heavy
elements, not hydrogen. Thus, this high column density might indicate
that W49B is embedded in an unusually metal rich environment.
More detailed investigation using higher resolution spectrometers
will be needed to resolve this question, and it is clearly
outside the scope of this work.

\subsection{W44}
\label{sec:w44}

\begin{figure}
\epsscale{0.35}
\plotone{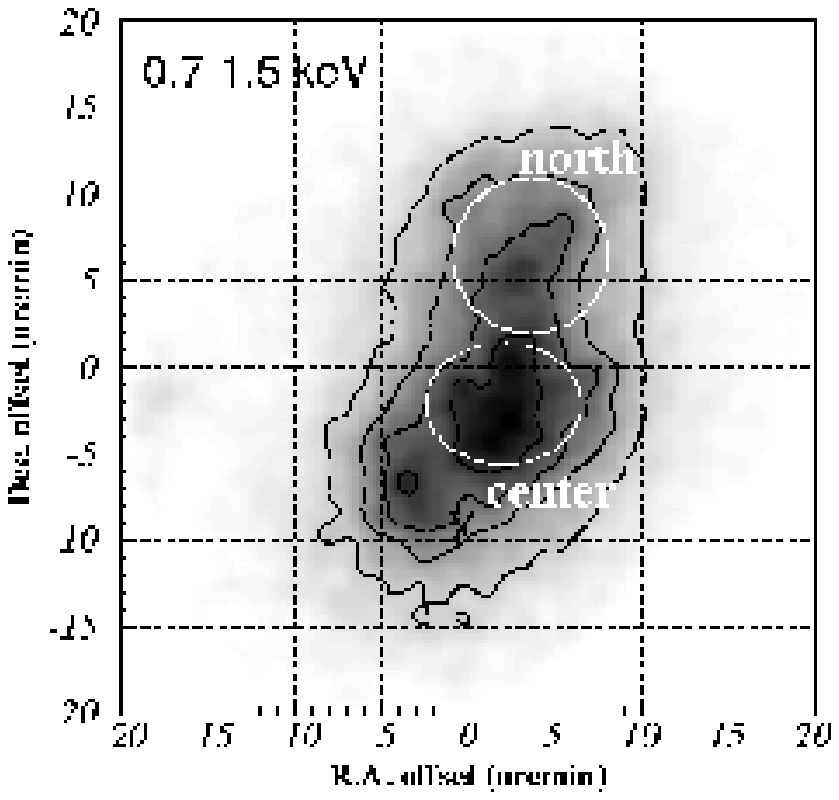}
\plotone{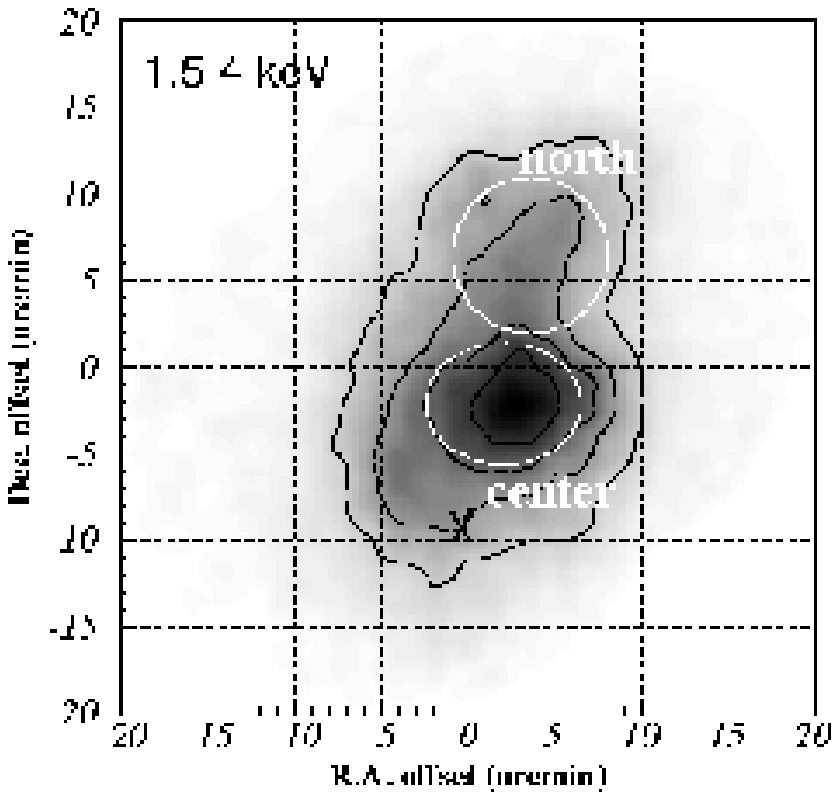}
\caption{Smoothed GIS images of W44 in the 0.7--1.5~keV and ({\it
left}) 1.5--4.0~keV ({\it right}) bands.  Grayscales are in linear
scale and contour levels in each figure are 20\%, 40\%, 60\%, and 80\%
of the peak surface brightness. Spectral extraction regions named
``center'' and ``north'' are shown in white ellipse and circle,
respectively.}
\label{w44img}
\end{figure}

We show in Figure~\ref{w44img} the smoothed images of W44 in the
0.7--1.5~keV and 1.5--4.0~keV bands using the Gas Imaging Spectrometer
(GIS).  The soft (0.7--1.5~keV) image has a larger angular extent than
the hard (1.5--4.0~keV) one, suggesting spectral variations due to a
temperature gradient.  Furthermore, clear differences are found in
fits to spectra from the central and northern regions shown in
Figure~\ref{w44img}.  The temperature of the central region,
0.84$^{+0.02}_{-0.05}$~keV, is higher than that of the northern,
0.64$^{+0.04}_{-0.05}$~keV. This is consistent with the fact that the
hard band image is more centrally concentrated. 
We also found higher
abundances of Mg and Si, significant at the 90\% confidence level, in
the central region: e.g., the abundance of Si is
1.18$^{+0.24}_{-0.12}$ in the center while in the north it is
0.50$^{+0.10}_{-0.04}$. This enhancement could be due to residual
ejecta, as first detected in the Cygnus Loop \citep{miyata98}.The best
fit ionization timescale is $>$10$^{12}$~\ntunit\ for both regions.
Details are tabulated in Table~\ref{w44param}. The \Chandra\ data also
show the remnant's hot and metal-rich projected center with bright
knots \citep{shelton04a}.  Note that the SIS data exhibited a slight
energy shift, which are also shown in the \Chandra\ data
\citep{shelton04a}.

\begin{figure}
\epsscale{0.4}
\plotone{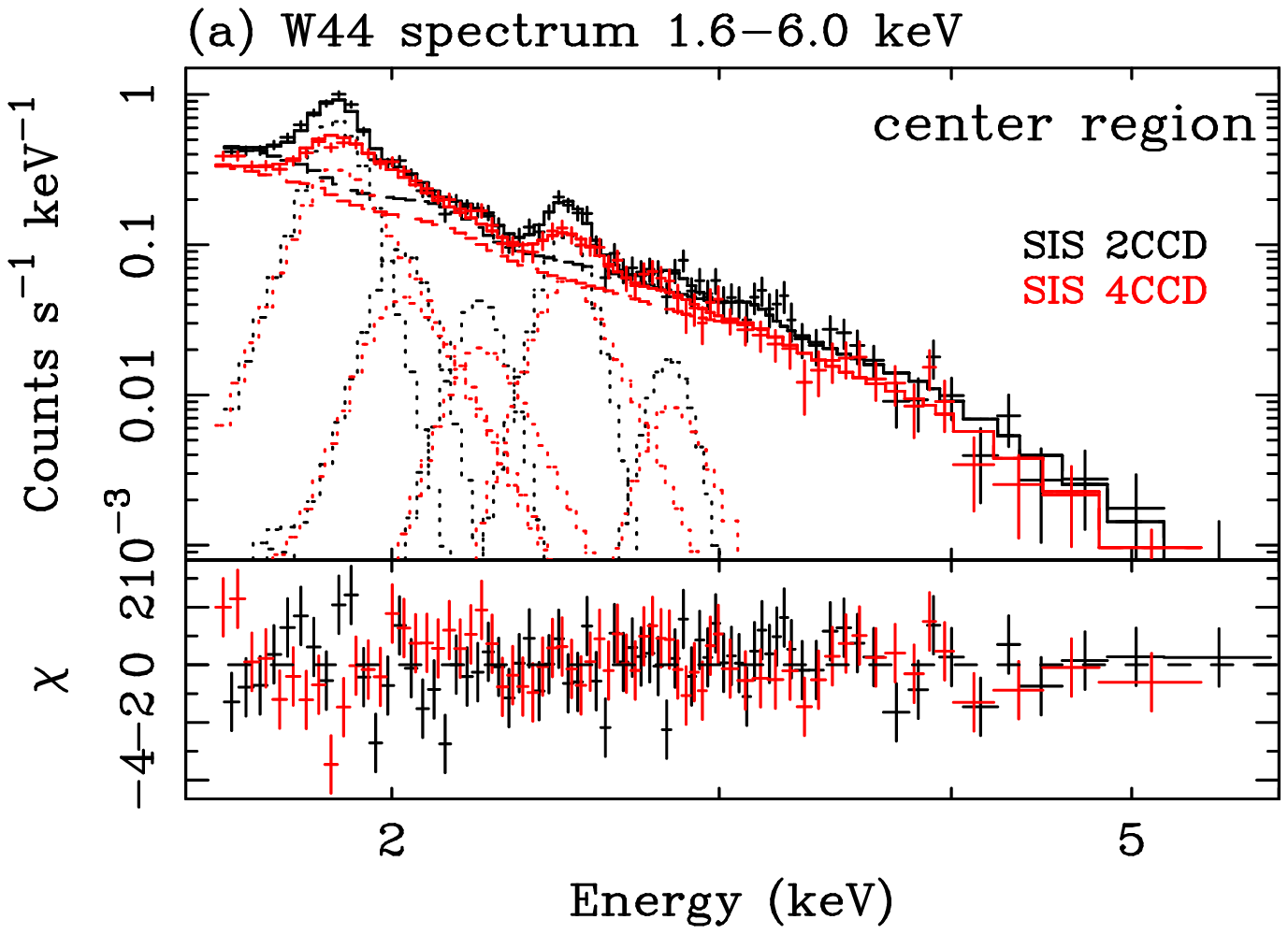}
\plotone{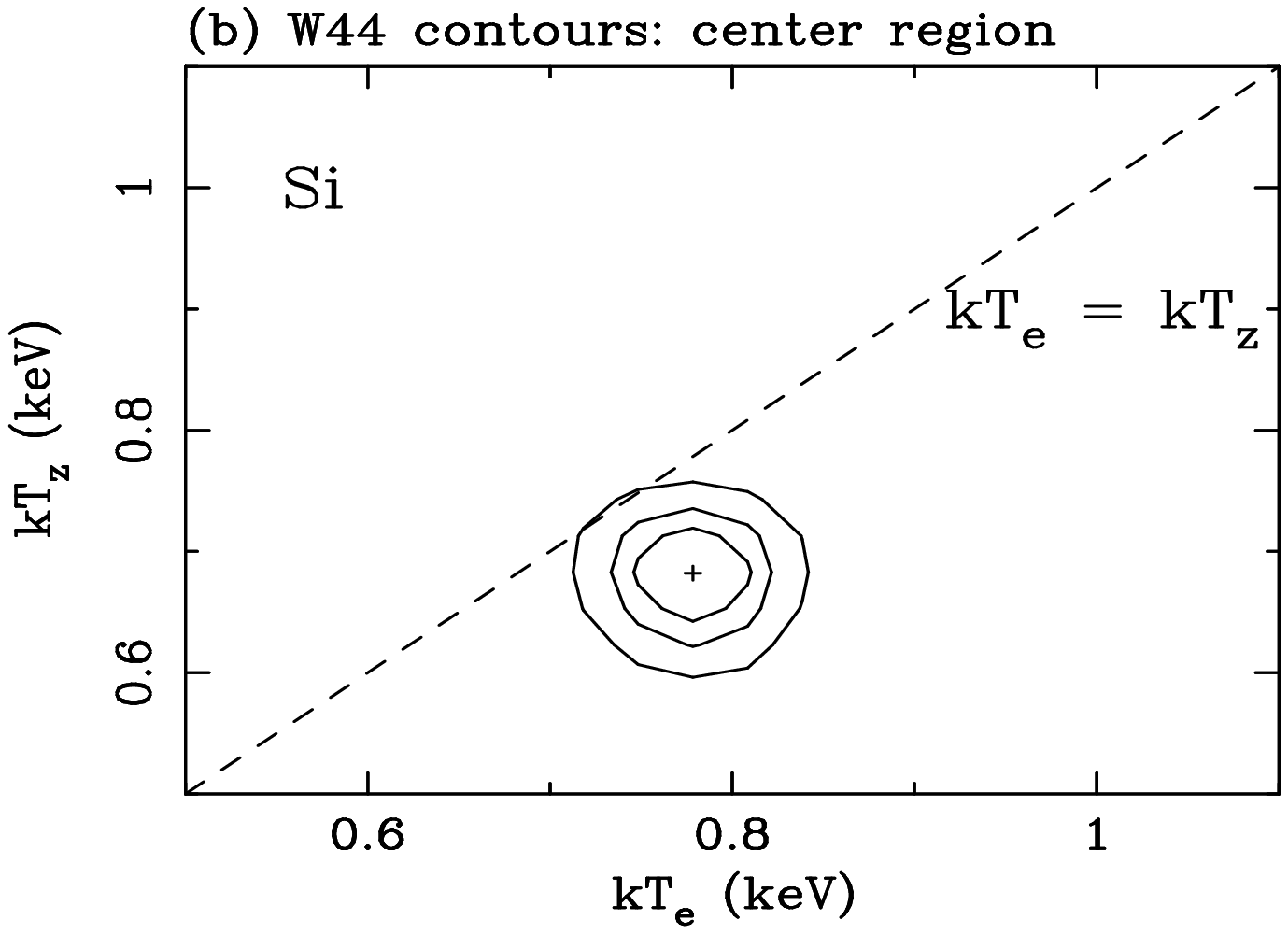}
\plotone{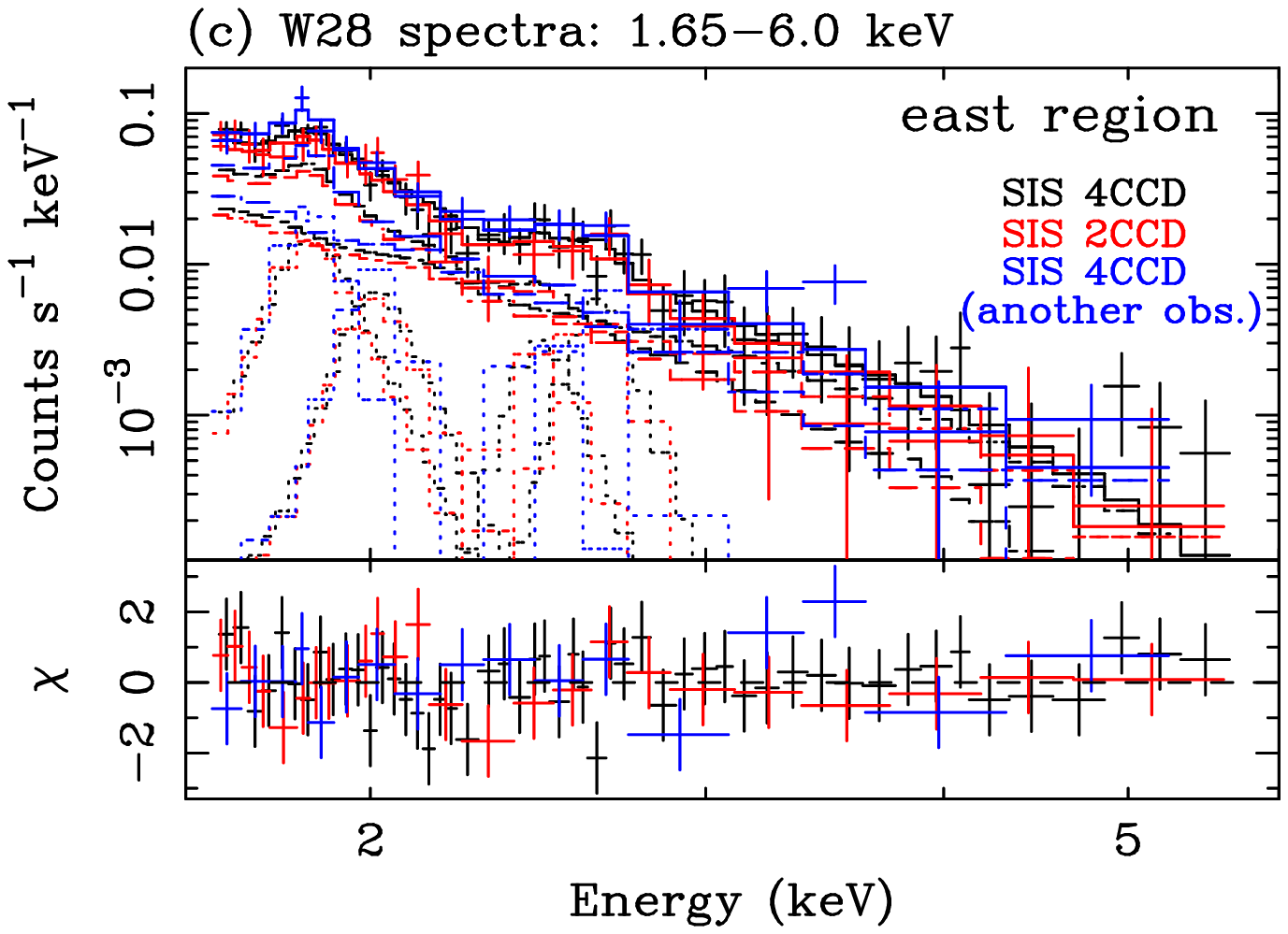}
\plotone{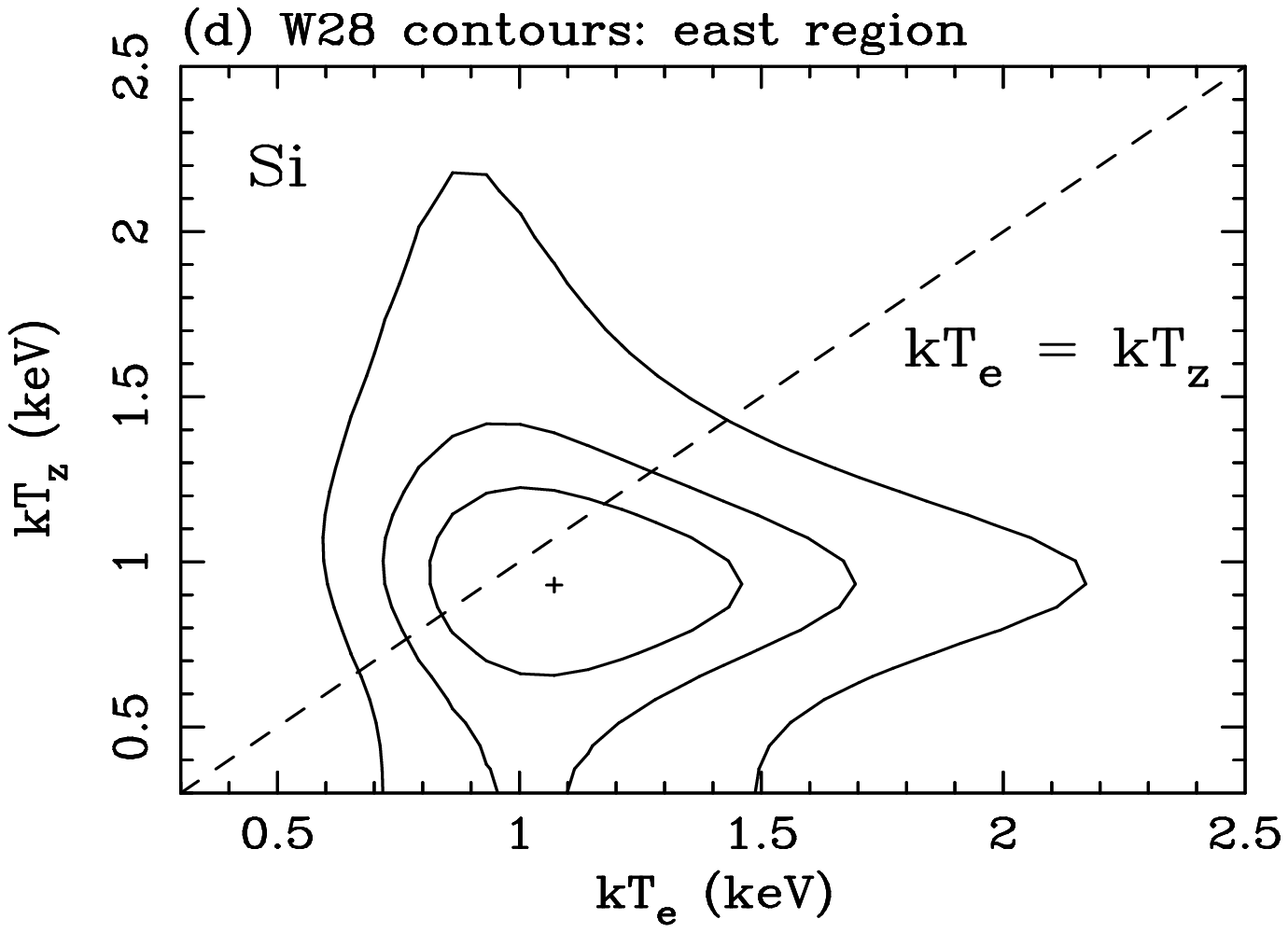}
\plotone{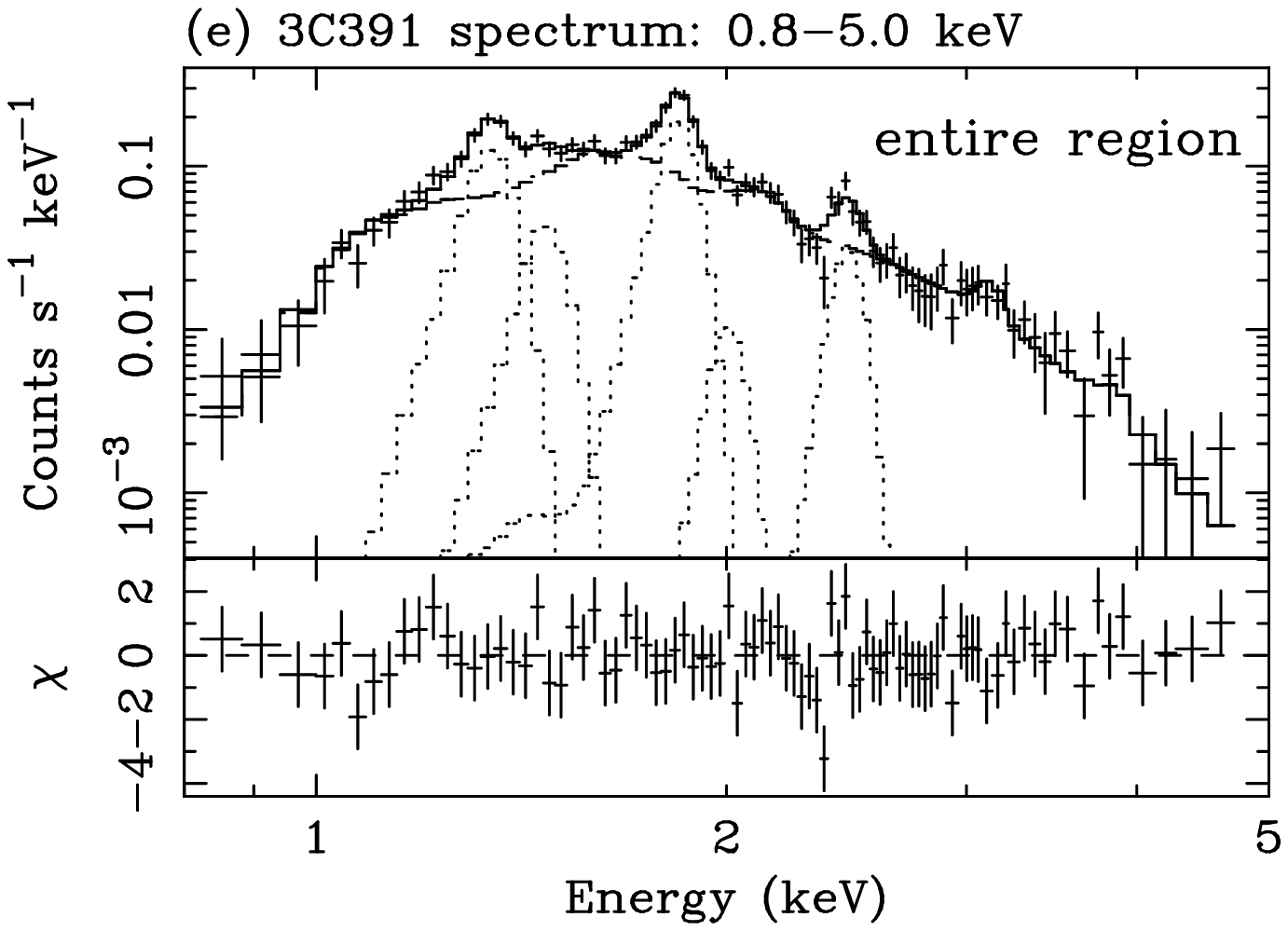}
\plotone{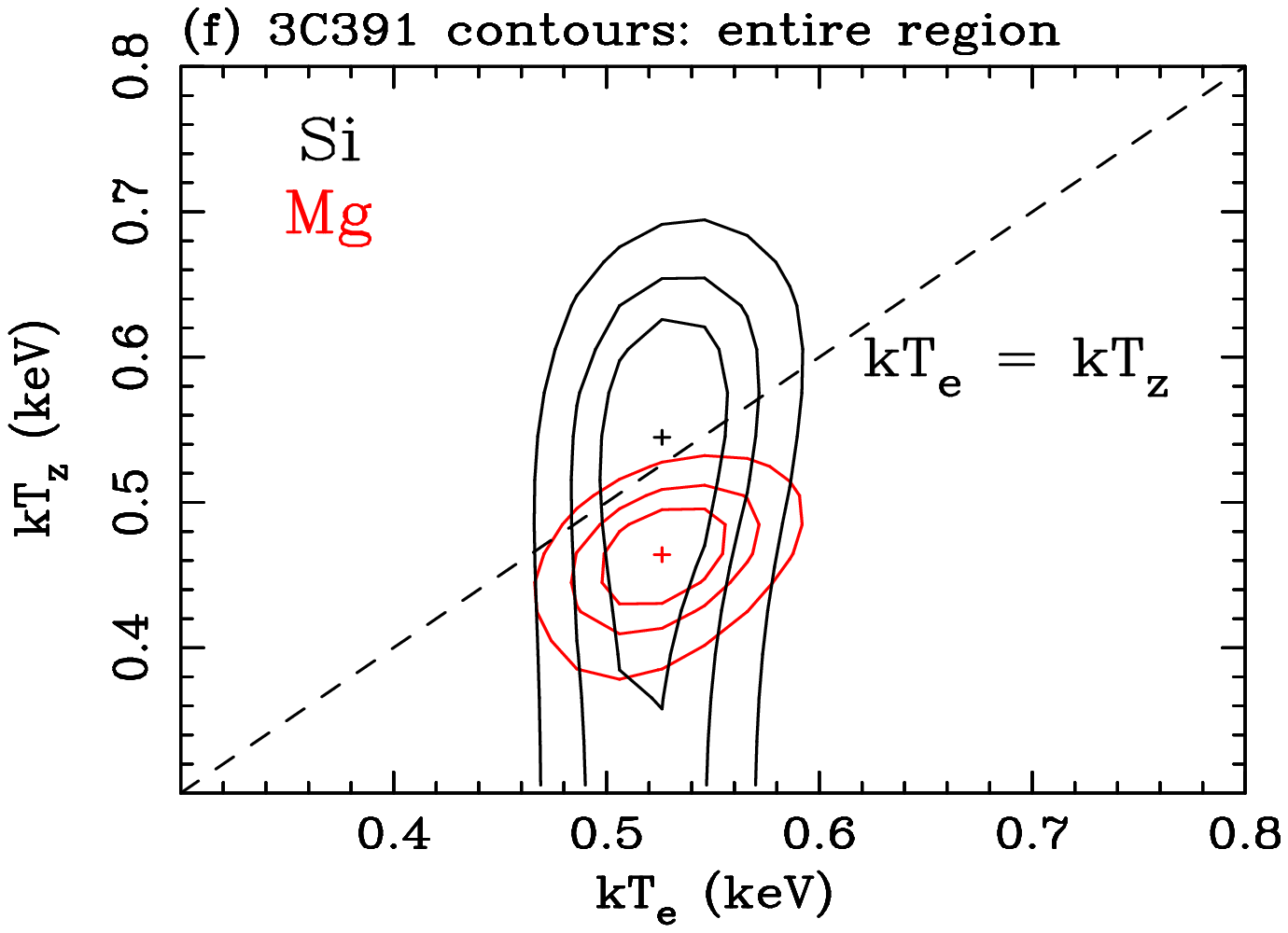}
\plotone{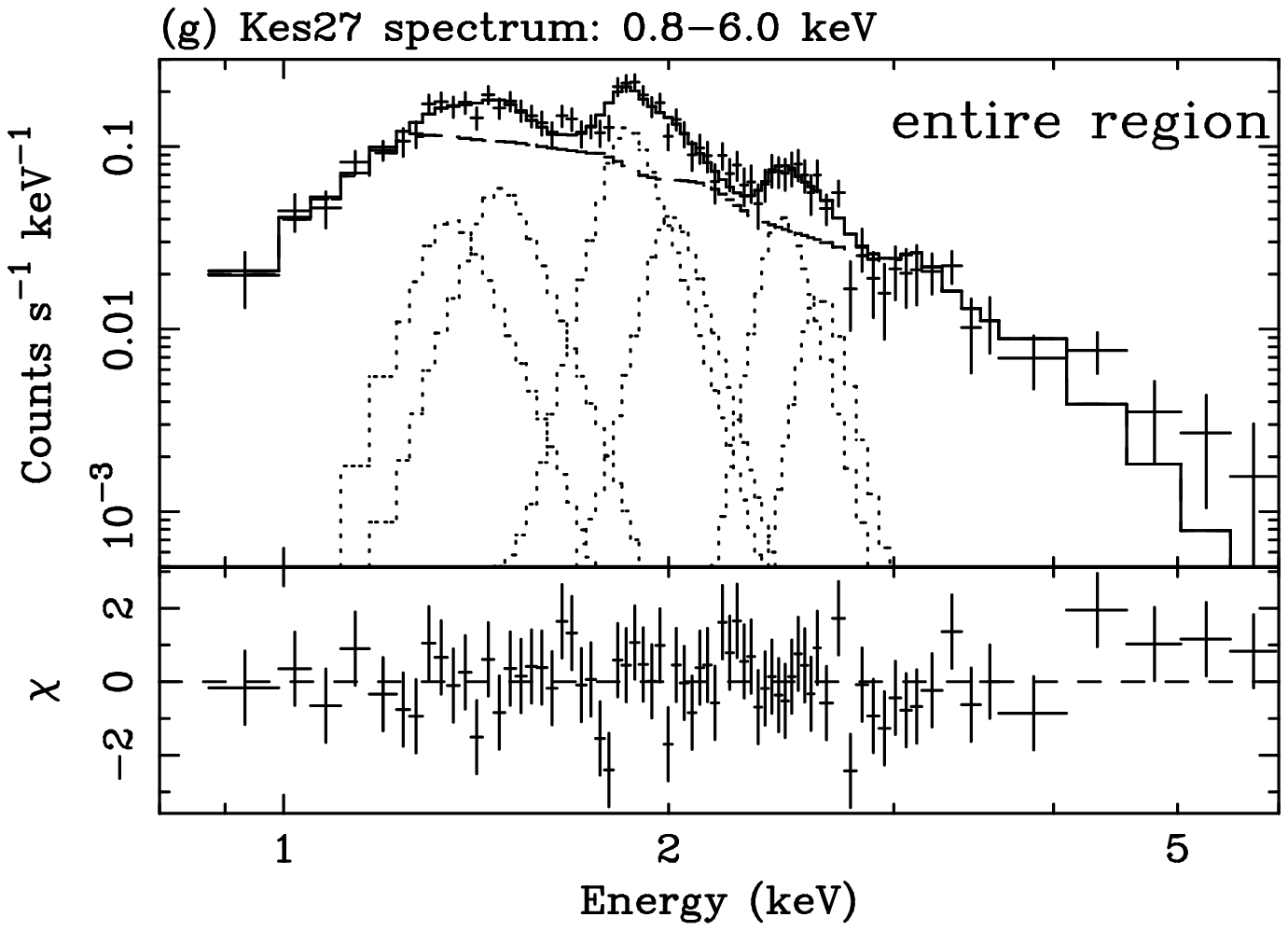}
\plotone{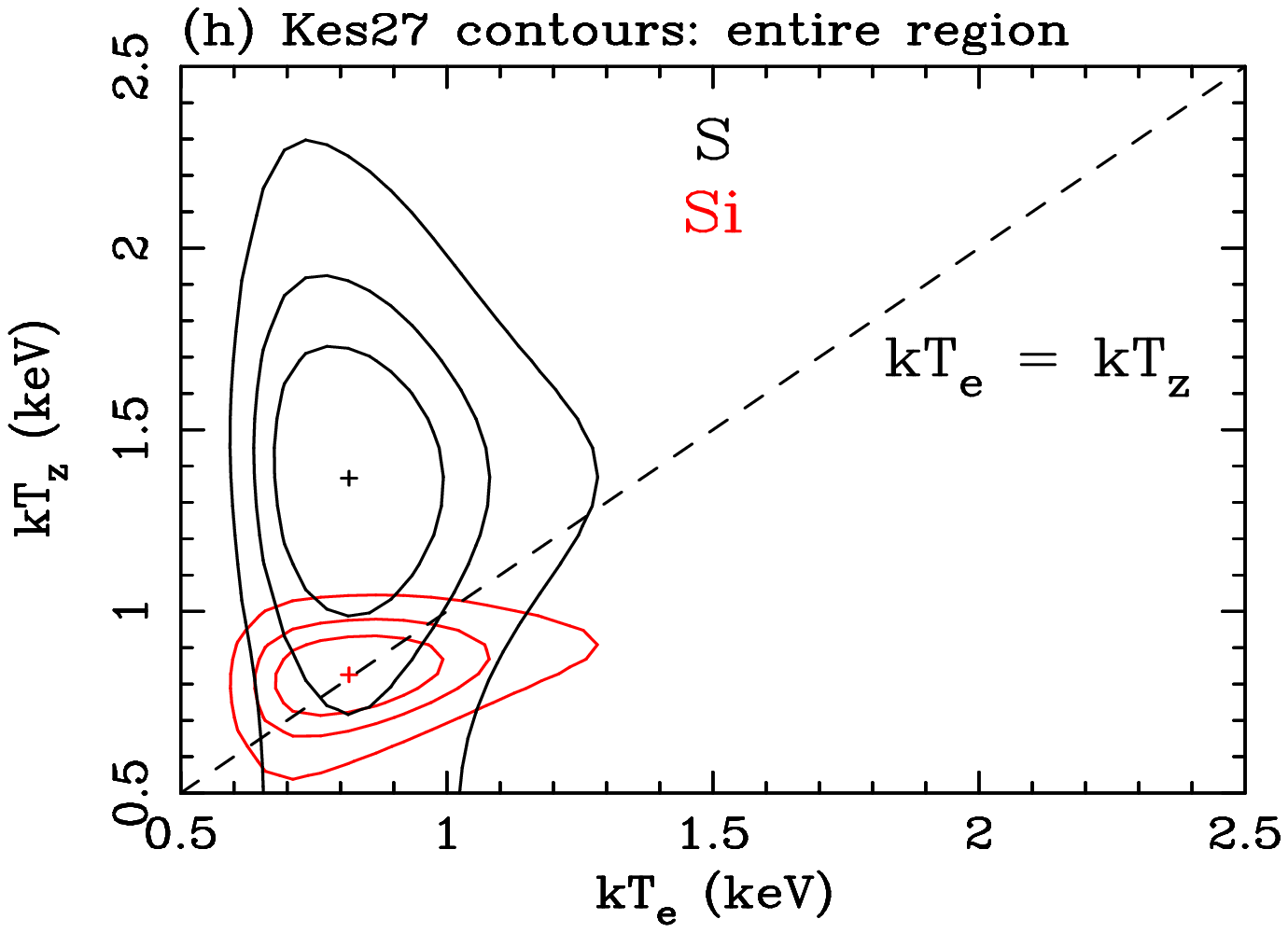}
\caption{X-ray spectra and confidence contours of the ratio of the ionization
temperature (\kTz) to the electron temperature (\kTe) for W44, W28,
3C391, and Kes~27. In each spectrum, the dashed line (and dash-dotted
line for W28) shows the continuum component while dotted lines
represent line features. The lower panel shows the residuals of the
fit.  In each contours, confidence levels are 99\%, 90\% and 67\%. The
dashed line represents \kTe\ = \kTz, meaning that the plasma is at the
CIE state.}
\label{cont}
\end{figure}

The ionization state of the W44 plasma was investigated using line
intensity ratios. In this fit, we used the 1.6--6.0~keV band spectra
(see Fig.~\ref{cont}(a)). An interstellar absorption value was fixed
to $8.9\times10^{21}$~cm$^{-2}$, which was obtained from the NEI fits.
The ionization temperature of Si in the central region, obtained from
the fitted \ratio\ ratio, is $\simeq0.7$~keV. This is
consistent with the electron temperature of $\simeq0.8$~keV as
shown in Figure~\ref{cont}(b), as well as with the NEI model
fitting. The inferred \kTz\ of Si in the north region is
$\sim0.8$~keV, with a large (0.5--1.5~keV) statistical uncertainty.

\subsection{W28}

The X-ray morphology of W28 is basically center-filled, but is more
complex than the other \mmsnrs.  It has ear-like segments of
limb-brightened shell in the northeast and
northwest \citep{rho02}.   In addition, we found with the \ASCA\ GIS
that the emission from the northeastern part is strong in the soft band
(0.5--1.5~keV) but not in the hard (1.5--4~keV).
This is also indicated by Figures~6 and 8 in \citet{rho02}.

We extracted the spectra from the central area, named ``center'' (a
rectangular region with 8\arcmin$\times$7\arcmin\ centered at
R.A.=18\fh00\fm26\fs, Dec.=$-$23\arcdeg26\arcmin49\arcsec) and the
eastern area, named ``east'' (an elliptical region with
4.8\arcmin$\times$4.1\arcmin\ centered at R.A.=18\fh00\fm36\fs,
Dec.=$-$23\arcdeg23\arcmin45\arcsec).  The spectrum was well fitted by
the two NEI ({\tt VNEI}) components with temperatures of 0.6~keV and
1.4~keV in each region (see Tab.~\ref{w28param}).  The ionization
timescale of both components are around $10^{12}$~\ntunit.  The
abundances of Ne, Mg, Si, S, which were linked together for the two
plasmas, are fitted with $\lesssim$1 times solar values
\citep{anders89}.  The column density is 5.1$\times10^{21}$~cm$^{-2}$
and 5.4$\times10^{21}$~cm$^{-2}$ for the center and the east,
respectively.  We found a larger fraction of high-temperature plasma
in the center region (the emission measure ratio is
0.51$^{+0.25}_{-0.19}$) than that in the east (the ratio is
0.31$^{+0.06}_{-0.04}$).  In contrast, the northeastern shell consists
of only a low-temperature plasma (\kTe=0.56~keV,
$n_{\mathrm{e}}t$=1.7$\times$10$^{13}$~\ntunit)
\citep{rho02}. Therefore, we suggest that low-temperature component
becomes dominant toward the northeastern rim of the remnant where
dense molecular clouds are present.  These results are consistent with
those of \ic\ \citep{kawasaki02}, implying that the plasma structure
of a hot interior and a cool exterior is present in W28.  On the other
hand, the southwestern shell region shows hard X-ray
emission. \citet{rho02} showed that the spectrum in this region can be
fitted with a temperature of 1.5~keV and $n_{\mathrm{e}}
t$=1.5$\times$10$^{11}$~\ntunit. This implies that the plasma
structure of W28 is more complex than allowed by the above simple
model.

The ionization state of the high-temperature plasma was examined
using line intensity ratios.  We used the spectra in the 1.65--6.0~keV
energy band although they include contamination by the
low-temperature plasma (see Fig.~\ref{cont}(c)).  We thus modeled the
column density and low-temperature plasma with all parameters fixed to
the values estimated from the component fits.  The ionization
temperature of Si in the east region, from the fitted \ratio\ ratio,
is close to the fitted electron temperature as shown in
Figure~\ref{cont}(d), although the poor statistics do not constrain
the temperatures well.  The inferred \kTe\ and \kTz\ of Si in the
central region are 1.19$^{+0.22}_{-0.18}$~keV and $>$1.06~keV,
respectively.  These results indicate that the high-temperature plasma
is near CIE in each region.  Hence, our direct estimate of the
ionization state suggests that the plasma in W28 has reached
ionization equilibrium, consistent with the results of the
NEI fits.

\subsection{3C391}

3C391 shows an elliptical X-ray morphology, with a size of
7\arcmin$\times$5\arcmin, and major axis oriented
from northwest to southeast.  \citet{chen01}
found that the \ASCA\ hard band (2.6--10~keV) image shows stronger
emission in the northwest, in anti-correlation with the soft
band (0.5--2.6~keV) image.  This difference is due to the larger
column density in the northwest.  The variation of the
column density across the remnant is in agreement with the presence of
a molecular cloud to the northwest. No temperature gradient has
been detected across the remnant, a result confirmed by 
\Chandra\ \citep{chen04}.

We extracted the spectrum of the entire remnant, in the energy band
0.8--5.0~keV.  We first applied a single NEI model, and obtained
an electron temperature and ionization timescale 
0.53$^{+0.4}_{-0.3}$~keV and 2.5 ($>$0.9) $\times10^{12}$~\ntunit.
These
suggest that the plasma has reached ionization equilibrium (see
Tab.~\ref{3c391param}).  We then measured the ionization degree of the
plasma using line intensity ratios (see Fig.~\ref{cont}(e)). Since the
spectrum shows prominent emission from \Hea\ lines of Mg, Si, and S
with good energy resolution, we left their center energies free; those
of the others were fixed to the theoretical values.
Figure~\ref{cont}(f) shows confidence contours of \kTz\ to \kTe\ for
Mg and Si, and indicates that the ionization temperature is comparable
with the electron temperature.  Thus, both our direct estimate and NEI
model fitting suggest that the plasma in 3C~391 is near ionization
equilibrium.

\subsection{\kes}
\label{kes27}

\kes\ shows diffuse, thermal X-ray emission with several unresolved
point-like sources \citep{seward96}.  Using \ASCA\ data,
\citet{enoguchi02} found a temperature gradient from the rim
(0.59$^{+0.04}_{-0.06}$~keV) to the interior (0.84$\pm$0.08~keV). This
temperature variation is similar to that in W44 (see
\S\ref{sec:w44}).

The 0.8--6.0~keV SIS spectrum of the entire remnant is well fitted by
a single NEI model (see Tab.~\ref{kes27param}).  The inferred electron
temperature, column density, and ionization timescale are
0.9$\pm$0.1~keV, 2.2$\pm$0.1$\times10^{22}$~cm$^{-2}$, and
$>3\times10^{11}$~\ntunit, respectively.  We measured the ionization
state of \kes\ using the line intensity ratios of Si and S, as shown
in Figure~\ref{cont}(g) and (h).  The center energies of all the lines
are fixed to the theoretical values for a 0.9~keV CIE plasma.  The
inferred \ratio\ ratio for Si indicates that the ionization
temperature of Si is close to the electron temperature. However, that
of S may require an ionization temperature larger than the continuum
temperature, suggestive of an overionized plasma, although this
statement is significant at a confidence level of $<$90\%.  The
results indicate that the plasma in Kes~27 is basically in ionization
equilibrium, but the strong S \Lya\ emission suggests possible
overionization.

\section{Discussion}
\label{discuss}
\subsection{Comparison with Shell-like SNRs}

\subsubsection{Plasma Structure}
We first summarize in Table~\ref{param_mmsnr} the spatial
distributions of electron temperature and density for the six
\mmsnrs.\@ \ic\ and W49B show clear evidence of temperature and
density gradients; they obviously indicate that these SNRs have a hot
interior surrounded by a dense, cooler exterior plasma, qualitatively
consistent with that of a standard shell-like SNR. We note that these
temperature and density gradients are not as large as those predicted
by the Sedov solution. Additionally, W44, W28, and Kes~27, show
temperature gradients; they have higher electron temperature towards
the center where the X-ray surface brightness is highest than towards
the edge.  A similar temperature gradient has been detected from
an ISM dominated shell-like SNR, the Cygnus Loop
\citep{miyata94,miyata98}, as tabulated in Table~\ref{param_mmsnr}.  
These results suggest that \mmsnrs\ have experienced
similar evolution to shell-like SNRs.  3C391 \citep{chen01,chen04} is
the only SNR with no detectable temperature or density gradient of
the six studied here.

\subsubsection{Ionization States}

We summarize in Table~\ref{ion_mmsnr} the electron and the
ionization temperatures of the \mmsnrs\ we have analyzed. Two of them,
\ic\ and W49B, show significantly higher ionization temperatures than
electron temperatures, suggesting the plasma is overionized. For the
other remnants, the two temperatures are comparable, suggesting that
the plasma has almost reached CIE.  This is supported by the long
ionization timescales obtained by the NEI model fits.

For comparison, we tabulate in Table~\ref{ion_shellsnr} the electron
temperature (\kTe), ionization timescale ($n_\mathrm{e}t$), and
ionization temperature (\kTz) of prominent shell-like SNRs. Cassiopeia~A
(Cas~A), Kepler's SNR, and Tycho's SNR exemplify the young
($\sim500$~yr) SNRs, and the others are examples of
middle-aged ($\gtrsim1000$~yr) SNRs.  We inferred
\kTz\ of each SNR by the following method: from \kTe\ and
$n_\mathrm{e}t$, the emissivities of \Hea\ and \Lya\ lines for heavy
elements are calculated using an NEI plasma code \citep{masai84,
masai94}; their ratio, \ratio, yields the ionization
temperature \kTz.  We estimated the \ratio\ ratios for Ne, Mg, Si, and
S using this method; the inferred \kTz\ values are consistent with one
another within $\simeq$0.2~keV. The validity of this estimate is
checked with the Cas~A spectrum as follows. We fitted the SIS spectrum
of Cas~A with two NEI ({\tt VNEI}) models; the measured \kTe\ and
$\log (n_\mathrm{e}t)$ of the high-temperature plasma are 3.02~keV and
10.83, respectively, from which we estimate \kTz\ of S as
$1.07^{+0.02}_{-0.04}$~keV.
On the other hand, we derived the \kTz\ of S for the high-temperature
plasma to be $0.94\pm0.07$~keV, from the fitted intensity ratio of
\Lya\ to \Hea\ (\ratio) as we did for \mmsnrs.
In this fit, the low-temperature plasma was modeled as another plasma
component, in the same way as for W28.  The
ionization temperatures derived with the two methods are consistent
within 0.1--0.2~keV. Thus, we confirmed that the
values tabulated in Table~\ref{ion_shellsnr} properly represent
\kTz\ of these shell-like SNRs, and can be compared with those
for \mmsnrs.

\begin{figure}
\epsscale{0.50}
\plotone{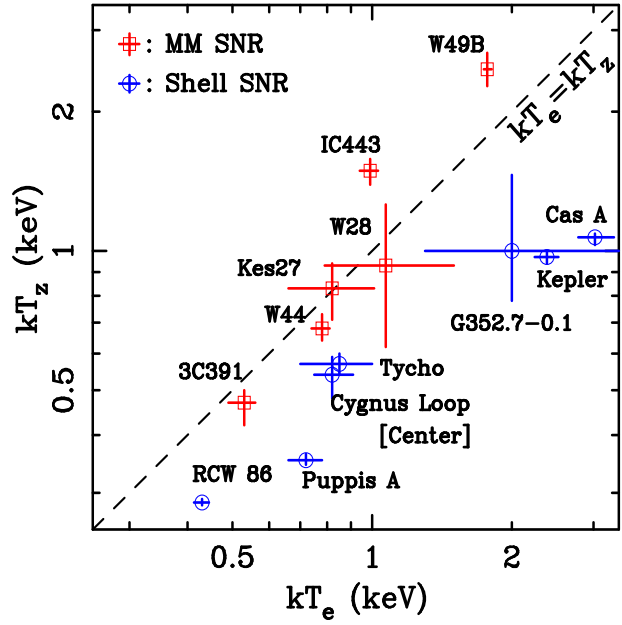}
\caption{The ionization temperature vs.\ the electron temperature for
various \mmsnrs\ (red open square) and Shell-like SNRs (blue open
circle). The dashed-line represents \kTe=\kTz, implying that the
plasma is at the CIE state.}
\label{ionize}
\end{figure}

Figure~\ref{ionize} shows the correlation between \kTe\ and \kTz\ for
both the MM and shell-like SNRs. We can see a systematic difference
between the two types of SNRs. All \mmsnrs\ show
\kTz~$\gtrsim$~\kTe, which means that the plasma has
reached CIE or a state of overionization.  On the other hand, all of the
shell-like SNRs show \kTz~$<$~\kTe, indicating underionized plasma conditions.
These results strongly suggest that \mmsnrs\ are systematically more
ionized than shell-like SNRs.

\subsection{Formation Mechanisms of the Overionized Plasma}
\label{form_plasma}

We also found that \Lya\ line emission from \ic\ and W49B is unusually
strong compared with \Hea, indicating that the interior
plasma in these remnants is overionized.  In the low density plasma
found in SNRs, we are accustomed to encountering
underionized plasma, like that detected in shell-like SNRs, because 
the ionization
timescale is typically long compared with the time since the bulk of
the gas was shock-heated.  Nevertheless, conditions can exist in which
such SNR plasmas can become overionized.

We consider two possible causes for overionization:
photoionization, and rapid gas cooling compared with recombination.  For
photoionization, a low-density SNR plasma itself does not produce
sufficient radiation to photoionize heavy elements, and there is no
strong X-ray source in any of these remnants (see Fig~\ref{w49bimg} for W49B,
and \citet{kawasaki02} for \ic.)  On the other hand, there are three
possible cooling mechanisms in a gas: radiation, expansion, and
thermal conduction. \citet{kawasaki02} compared these timescales with
that of recombination, and showed that in \ic\ thermal conduction
could drive the plasma into an overionized state.
A similar comparison can be carried out for W49B.
The conduction
timescale,  $9\times10^{10}(n_1/2.5~\mathrm{cm}^{-3})
(l_{\mathrm{T}}/10^{19}~\mathrm{cm})^2 (kT_1/1.8~\mathrm{keV})^{-5/2}
(\ln \Lambda/32.2)$~sec, is shorter than the recombination timescale
of $4\times10^{11} (n_1/2.5~\mathrm{cm}^{-3})$~sec, while the radiative
cooling timescale is $3\times10^{14} (kT_1/1.8~\mathrm{keV})
(\theta_1/1.7\arcmin)^{3/2} (d/11.4~\mathrm{kpc})^{1/2}$~sec. Here,
$n_1$, $\theta_1$, and $kT_1$ are the number density, the radius in
arcminutes, and the temperature of the inner plasma of W49B, respectively,
and $l_{\mathrm{T}}$, and $\ln \Lambda$ are the scale length of the
temperature gradient and Coulomb logarithm. We note that the cooling
timescale via expansion is 12.5 times the remnant's age
\citep{masai94}, meaning that the remnant has not been cooled enough
at present time.  This indicates that thermal conduction is the
dominant process producing the overionized plasma in W49B as well as
in \ic.

The presence of a magnetic field will reduce the heat flux in
principle, because diffusion perpendicular to the field direction is
suppressed. The magnetic field reduces the heat flux by a factor of
$\sim\langle\cos^2\theta\rangle = 1/3$ if it is tangled, but the
conduction timescale still remains comparable to the
recombination timescale.  Although detailed information about the
magnetic field in W49B is unavailable, we conclude from the above
argument that it is possible for the gas to be overionized due to
thermal conduction. The fact that we have not detected in any SNR as
large a temperature distribution as predicted by the Sedov solution
further suggests that thermal conduction is operating in many
remnants.

\begin{figure}
\epsscale{1.0}
\plottwo{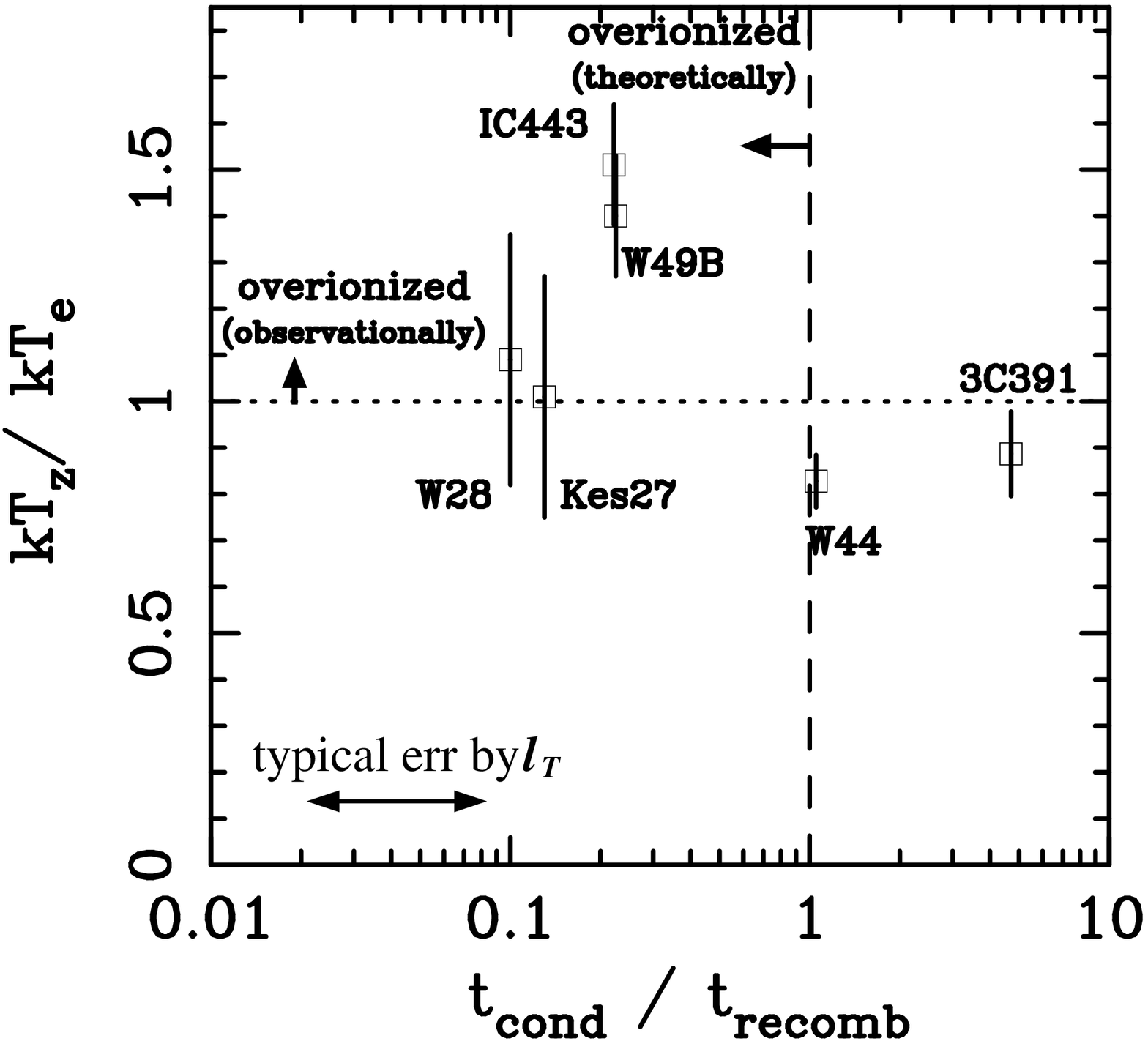}{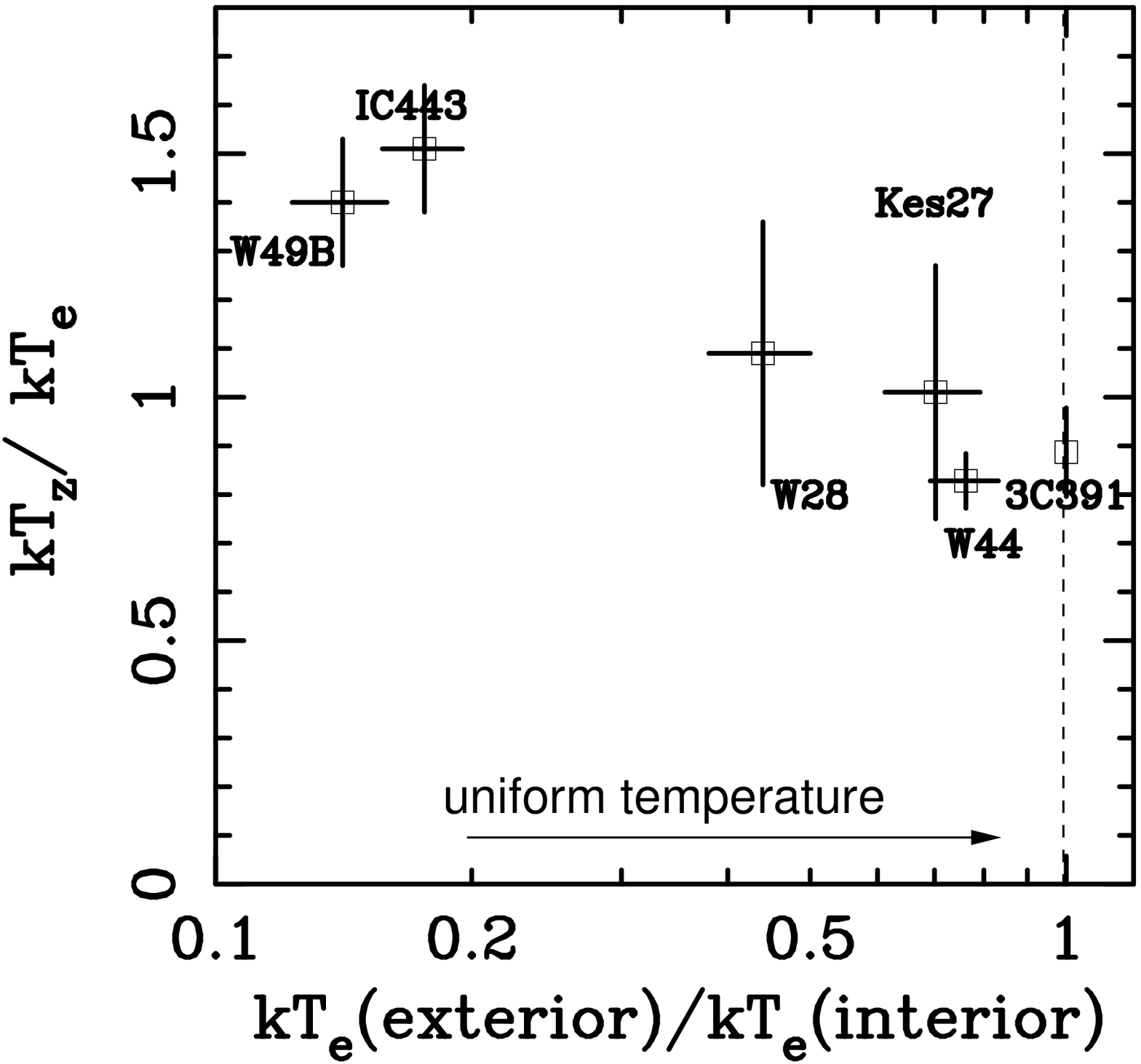}
\caption{Comparisons between overionized SNRs and the other \mmsnrs.
{\it left}: The relations of the ionization states (\kTz/\kTe) to the
timescale ratio ($t_{\mathrm{cond}}/t_{\mathrm{recomb}}$) for the
\mmsnrs. The typical error of the ratio due to uncertainties in $l_T$
is also shown at the lower left of the panel.  The condition
$t_{\mathrm{cond}}/t_{\mathrm{recomb}} > 1$ means that it is hard for
the plasma to be overionized due to thermal conduction. {\it right}:
The relations of the ionization states (\kTz/\kTe) to the electron
temperature gradients (\kTe\ (exterior)/\kTe\ (interior)) for the
\mmsnrs. The dashed line denotes \kTe\ (exterior)/\kTe\ (interior) =
1, meaning a uniform temperature within the remnant. For 3C391 with no
detection of temperature difference within the remnant, we assumed
the uniform temperature.}
\label{comp_ion}
\end{figure}

The process of thermal conduction should arise in all SNRs, as
\citet{cox99, shelton99b} have introduced and discussed thermal
conduction over the lifetime of an SNR.  Therefore, we investigated
the differences between overionized SNRs and the other \mmsnrs\ by
comparing the ionization states to the parameters related to thermal
conduction.  Figure~\ref{comp_ion} ({\it left}) show the ionization
states (i.e., \kTz/\kTe) plotted against the ratios between the
conduction timescale and the recombination timescale
($t_{\mathrm{cond}}/t_{\mathrm{recomb}}$) for the six remnants we have
analyzed. In the calculation of the conduction timescale, we ignored
the effect of the magnetic field.  The ratio of the timescales has
strong dependence on the density, temperature gradient, and electron
temperature as $t_{\mathrm{cond}}/t_{\mathrm{recomb}} \propto n^2 \>
l_T^2 \> (kT_{\mathrm{e}})^{-5/2}$.  This figure suggests that 
conduction in W44 and 3C391 is not strong enough to produce overionization,
consistent with the observations.  On the other
hand, our estimates suggest Kes~27 and W28 might be overionized, 
even though our analysis reveals no clear evidence.  Two possibilities are
considered for the discrepancy: one is that we could not detect the
evidence because of the poor statistics as described in \S\ref{kes27}
for Kes~27; the other is that the magnetic field or some other
process reduces the heat flux.  We also compared the ionization
state to the temperature gradient within the remnants because the
temperature gradient strongly affects the thermal conduction as
described above.  In Figure~\ref{comp_ion} ({\it right}) we compare the
ionization state against the temperature ratio between the exterior
and the interior (\kTe\ (exterior)/\kTe\ (interior)) for the six
\mmsnrs.  This figure indicates that overionized SNRs have a larger
temperature gradient than the other \mmsnrs (Note, however that
temperatures of the interior and the exterior were not estimated in
the same way for the six remnants.)  We also suggest that \mmsnrs\
with no overionization, such as W44 and 3C391, can be those which
conduction has taken further from an overionized state since thermal
conduction gradually reduces the temperature gradient within the
remnant.

\subsection{SNR Evolution and Formation of \mmsnr }

\begin{figure}
\epsscale{0.8}
\plotone{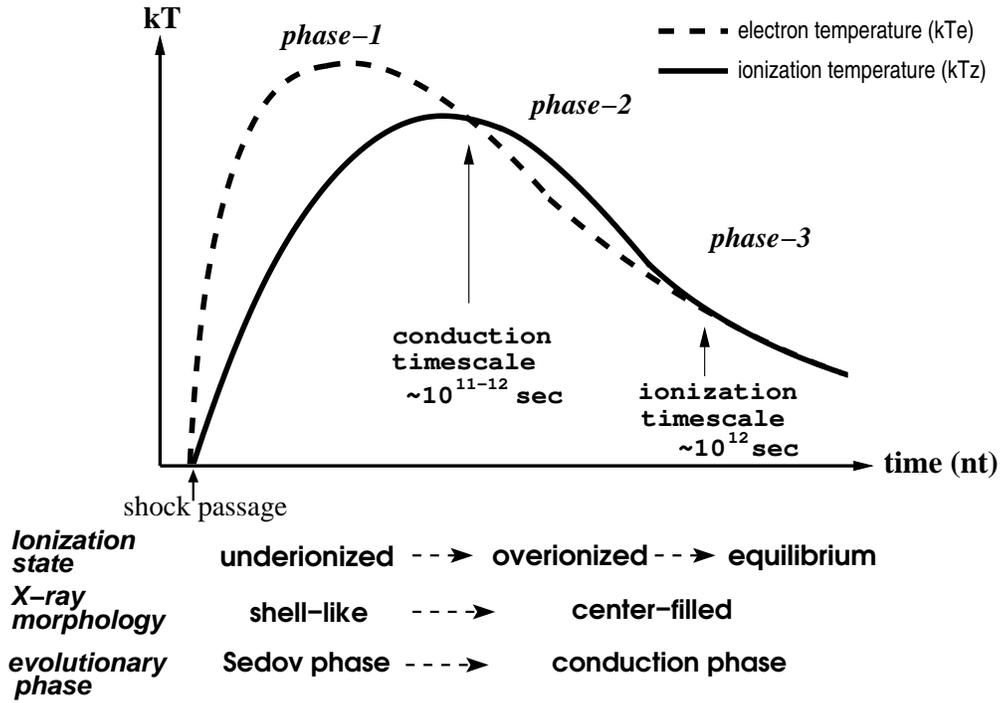}
\caption{Schematic view of the evolution of the ionization state in an
SNR along time.  The horizontal line indicates the time normalized by
the plasma density, and the vertical line qualitatively shows the
electron and ionization temperatures. The ionization temperature
follows the electron temperature with a delay of $n_{\mathrm{e}} t
\sim 10^{12}$~\ntunit.}
\label{schematic_ion}
\end{figure}

As we described in \S\ref{form_plasma}, thermal conduction arises in
an SNR, and it equilibrates the temperatures while keeping the
pressure balanced. It suggests that the hot interior plasma gradually
becomes denser.  Therefore, we propose the following unified scenario
for SNR evolution.  We present in Figure~\ref{schematic_ion} a
schematic view on the evolution of the ionization state in a remnant.
Numerical simulations by \citet{shelton98,shelton99a} also predict
both the center-filled X-ray morphology and overionization as a result
of the SNR evolution.

In a young remnant at an early adiabatic or Sedov phase, the strong
front shock heats the ambient ISM to high temperatures and bright
X-rays appear along the outer rim.  The hot gas within the post-shock
interior is rarefied compared to the shell regions, and hence, the
``shell-like'' morphology in X-rays is formed.  The observed plasma is
mainly from recently shocked regions where ions are being ionized to
higher ionization states by electrons on the timescale of
$n_{\mathrm{e}} t \sim 10^{12}$~\ntunit.  Therefore, an underionized
plasma is detected ({\it phase-1} in Figure~\ref{schematic_ion}).  

As the remnant evolves, the hot interior plasma gradually approaches a
uniform temperature via thermal conduction, while maintaining the
pressure equilibrium.  Thus, the interior also becomes denser,
resulting in an enhancement of the X-ray emission from the central
region. In contrast, the shock speed decreases and the X-ray emission
from the shock-heated shell becomes too soft to pass through the
ISM.\@ These lead the X-ray morphology to change from ``shell-like''
to ``center-filled'' on the conduction timescale, which is the
formation mechanism of \mmsnr. \@ Since the conduction timescale is
roughly comparable to the ionization timescale, $t_{\mathrm{cond}}
\sim t_{\mathrm{recomb}}\simeq10^{12}$~\ntunit, ionization states of
the remnants reach nearly an ionization equilibrium as derived from
our analysis of \mmsnrs.\@ This phase can be termed the ``conduction
phase'' since thermal conduction plays an important role to the SNR
plasma.

If the conduction timescale is shorter than the ionization timescale,
the interior plasma could become ``overionized'' over a certain
interval, as observed in \ic\ and W49B ({\it phase-2} in
Figure~\ref{schematic_ion}).  It is uncertain how long this interval
lasts, but the plasma will eventually reach another ionization
equilibrium because conduction becomes insufficient as the interior
plasma cools and the temperature gradient disappears ({\it phase-3} in
Figure~\ref{schematic_ion}).  When the conduction becomes fully
efficient, no temperature gradient remains in the remnant, as observed
in 3C391. Figure~\ref{comp_ion} ({\it right}) may indicate the
evolution from the overionized SNR ({\it phase-2}) to the fully
conducted SNR ({\it phase-3}).

This interpretation is supported by the X-ray morphological analysis
of SNRs in the Magellanic Clouds that ``centrally brightened'' SNRs
are larger in size than ``shell-type'' SNRs \citep{williams99}, and by
G65.3+5.7, the remnant newly identified to be in transition from
shell-like to MM \citep{shelton04b}.  
It is controversial to include W49B to the \mmsnr\ type since it has
been considered to be ejecta-dominated plasma.  However, from the
point of view of ionization, thermal conduction plays an important
role in W49B, whether it is ejecta or not, and it supports the above
interpretation.

\section{Conclusions}
\label{conclude}

We have discovered the following results from the systematic analysis
of the six \mmsnrs.
\begin{enumerate}
\item Hot interior and cool outer plasma components are detected in five
MM remnants. A higher density is found in the cool
exterior than that in the hot interior from both \ic\ and W49B. These
results suggest a plasma structure consisting of a central hot
region surrounded
by a cooler, denser shell, similar to the expected
structure of a shell-like SNR.
\item The X-ray emitting gas in \ic\ and W49B is ``overionized;'' the
gas in the other \mmsnrs\ are found to be near ionization equilibrium.
As a consequence, the degree of ionization in
\mmsnrs\ is systematically higher than in shell-like SNRs.
\end{enumerate}
The first result indicates that \mmsnrs\ have experienced a similar
evolution to shell-like SNRs, and the second result indicates that
\mmsnrs\ are more evolved than shell-like SNRs. We suggest
that thermal conduction plays an important role in producing the
overionized interior plasma in \ic\ and W49B.  It also acts to diminish
the temperature and density gradients within all remnants, leading
naturally to a ``center-filled'' X-ray morphology. Hence, we conclude that
\mmsnrs\ are those which have evolved over a conduction timescale of
$\sim10^{11-12}$~sec with thermal conduction active within the
remnants. It is also suggested from the fact that plasmas of \mmsnrs\
are nearly in CIE, with the estimated ionization timescale of
$\sim10^{12}$~sec.  We propose to call this phase the ``conduction
phase'' since thermal conduction drastically changes the plasma
structure of an SNR in this phase far from that in the Sedov phase.
By introducing thermal conduction, we showed that \mmsnrs\ are not
special SNRs but those in a certain stage of an SNR
evolution.  This leads to the construction of a unified model of
SNR evolution.

\acknowledgments 
We thank Dr. Manabu Ishida and Dr. Kuniaki Masai for
helpful discussion, and Dr. Kazuo Makishima for thoughtful comments
and suggestions.




\clearpage

\begin{deluxetable}{lcccc}
\tablecaption{The list of \mmsnr\ candidates with
\ASCA\ observation logs \label{list}}
\tablehead{
SNR name & Size (arcmin) & Group\tablenotemark{a} & \ASCA\ Obs.
& Exposure time\tablenotemark{b}}
\startdata
W28     & 42            & A     & 1994/4, 1995/9 & 50ks(1994/4, 4CCD) \\
W44     & 28$\times$34  & A     & 1994/4, 10 & 24ks(1994/4), 26ks(1994/10) \\
3C400.2 & 28            & A     & 1994/10\tablenotemark{c}, 1996/4 & \\
\kes    & 20            & A     & 1994/8 & 23ks\\
MSH~11-61A & 12         & A     & 1994/3\tablenotemark{d}, 1995/1 & \\
3C391   & 8             & A     & 1994/4 & 44ks \\
CTB~1   & 34            & A     & 1996/1 &\\
W51C    & 25$\times$19  & B     & 1995/10 &\\
CTA~1\tablenotemark{e}   & 80            & B     & 1996/1, 1997/2 &\\
W63     & 95$\times$65  & B     & 1997/6 & \\
HB21    & 120$\times$90 & B     & 1997/6 & \\
\ic     & 50            & C    & 1993/4, 1994/3, 1998/3, 1999/3 
					& 72ks(1993/4) \\
Kes~79  & 11            & C    & 1995/4 & \\
HB3     & 70            & C     & 1996/8 & \\
G327.1$-$1.1\tablenotemark{e} & 14       & D & 1996/3 & \\
CTB~104A & 80           & D     & no observation  & \\
W49B    & 3.5           & E     & 1993/4, 5, 10, 11 & 46ks(1993/4)\\
3C~397  & 6             & E & 1995/4 & \\
MSH~11-54 & 10          & F  & 1993/8 & \\
\enddata
\tablenotetext{a}{classification of the \mmsnr\ candidates assigned by
\citet{rho98}: The group A remnants are the prototypical \mmsnrs,
group B remnants are probably \mmsnrs, and group C remnants show
additional spatial components. Group D remnants are those whose data
quality are too poor to allow any determination. Group E are probably
not \mmsnrs; nevertheless, they have similar characteristics, and so
their classification remains uncertain. Group F is definetely not an
\mmsnr.}  
\tablenotetext{b}{Effective SIS (SIS0 + SIS1) exposure time after 
screening are listed for the targets that we analyzed in this paper.}
\tablenotetext{c}{The target was out of the field of view
because of a pointing error} 
\tablenotetext{d}{Due to observation at
small Earth angle, only a small fraction of the SIS data remains after
the screening \citep{slane02}.}  
\tablenotetext{e}{\ASCA\ observations revealed the non-thermal component 
that suggests these remnants are a true composite type \citep{slane97,sun99}.}

\end{deluxetable}

\begin{deluxetable}{lcccc}
\tablecaption{The best-fit parameters of the two CIE model for the W49B spectra\label{w49bparam} }
\tablehead{
 & parameter~(unit) & & value & (90\% c.r.) }
\startdata
wabs\tablenotemark{a} & \Nh~(10$^{22}$cm$^{-2}$) & & 
 5.23 & (5.13--5.34)\\ \hline
Low-T plasma & \kTe~(keV) & & 0.24 & (0.22--0.28)\\ 
 & E.M.\tablenotemark{b} & & 2.7 & (1.0--5.8)  \\ \hline
High-T plasma & \kTe~(keV) & & 1.70 & (1.66--1.72)\\
 & E.M.\tablenotemark{b} & & 0.094 & (0.090--0.097) \\ \hline
 & Mg & & 0.55 & (0.18--0.92) \\
 & Al & & 0 & ($<$1.74) \\
 & Si & & 3.5 & (3.2--3.8) \\
Abundance & S & & 3.9 & (3.7--4.3) \\
 & Ar & & 3.2 & (2.8--3.7) \\
 & Ca & & 4.5 & (4.0--5.1) \\
 & Fe & & 5.5 & (4.9--6.2) \\
 & Ni & & 16 & (7.4--25) \\ \hline 
\ion{Ar}{18} K$\alpha$
& flux\tablenotemark{c} & & 7.5 & (4.4--10.7)\\
\ion{Ca}{20} K$\beta$ 
& flux\tablenotemark{c} & & 5.2 & (3.4--7.1)\\
\ion{Cr}{23} K$\alpha$ 
& flux\tablenotemark{c} & & 2.0 & (0.7--3.4) \\
\ion{Mn}{24} K$\alpha$
&flux\tablenotemark{c} & & 1.0 & ($<$1.6) \\ \hline 
\multicolumn{2}{c}{$\chi^2$/d.o.f.} & & \multicolumn{2}{c}{233/146}
\enddata
\tablecomments{Each abundance is relative to the solar value and those of 
two plasma components are linked together. The 90\% confidence range
(c.r.) for temperature, abundances, and line parameters are given in
brackets. The line width of each Gaussian profile is fixed to 0.  }
\tablenotetext{a}{The column density of an interstellar 
absorption assuming solar abundances.}
\tablenotetext{b}{Emission Measure: the unit is 
4$\pi d^2 \times 10^{14}$~cm$^{-5}$ where $d$ is the distance to W49B.}
\tablenotetext{c}{The unit is 10$^{-5}$ photons~cm$^{-2}$~s$^{-1}$.}
\end{deluxetable}

\begin{deluxetable}{lccc}
\tablecaption{The best-fit parameters for the W44 spectra 
\label{w44param} }
\tablehead{
& & Center & North \\ 
\cline{3-4} 
parameter(unit) & & value (90\% c.r.) & value (90\% c.r.) \\
}
\startdata
~~\Nh~(10$^{22}$cm$^{-2}$) & & 
 0.89 (0.84--1.04) & 1.03 (0.90--1.25) \\ 
~~\kTe~(keV) & & 
0.84 (0.79--0.86) & 0.64 (0.59--0.68) \\ 
~~$n_{\mathrm{e}} t$~(10$^{12}$ cm$^{-3}$~s) & & 
10 ($>$1.1) & 9.5 ($>$1.0) \\
~~E.M.\tablenotemark{a} & & 
6.1 (5.4--7.8) & 10.5 (8.4--13.5) \\ \hline
Abundance & & & \\
~~Ne & & 0.60 (0.16--0.82) & 0.33 (0.16--0.58) \\
~~Mg & & 0.78 (0.51--0.89) & 0.21 (0.13--0.30) \\
~~Si & & 1.18 (0.96--1.32) & 0.50 (0.46--0.60) \\
~~S & & 0.82 (0.61--1.01) & 0.50 (0.33--0.65) \\
~~Fe & & 0 ($<$0.04) & 0.02 ($<$0.10) \\ 
~~Ni & & 0 ($<$0.2) & 0 ($<$0.4) \\ 
\hline 
$\chi^2$/d.o.f. & & 91/87 & 83/83 \\ 
\hline
\enddata
\tablecomments{Each abundance is relative to the solar value.
The 90\% confidence range (c.r.) for temperature, ionization
timescale, and abundances are given in brackets.}
\tablenotetext{a}{Emission Measure: the unit is 
4$\pi d^2 \times 10^{12}$~cm$^{-5}$ where $d$ is the distance to W44.}
\end{deluxetable}

\begin{deluxetable}{lccc}
\tablecaption{The best-fit parameters for the W28 spectra 
\label{w28param} }
\tablehead{
& & Center & East \\ 
\cline{3-4} 
parameter(unit) & & value (90\% c.r.) & value (90\% c.r.)
}
\startdata
~~$N_{\mathrm{H}}$\tablenotemark{a}~(10$^{21}$cm$^{-2}$) & & 
 5.1 (4.9--5.9) & 5.4 (5.1--6.1) \\ 
\hline
Low-T plasma ({\tt VNEI}) & & & \\
~~$kT$~(keV) & & 
0.62 (0.59--0.65) & 0.61 (0.60--0.63) \\ 
~~$n_{\mathrm{e}} t$~(10$^{12}$ cm$^{-3}$~s) & & 
31 ($>$2.7) & 15 ($>$1.0) \\
~~E.M.\tablenotemark{b} & & 
1.7 (1.2--2.2) & 0.9 (0.8--1.1) \\ 
High-T plasma ({\tt VNEI}) & & & \\
~~$kT$~(keV) & & 
1.41 (1.31--1.67) & 1.37 (1.31--1.49) \\
~~$n_{\mathrm{e}} t$~(10$^{12}$ cm$^{-3}$~s) & & 
0.5 (0.3--2.1) & 0.4 (0.3--0.8) \\
~~E.M. ratio\tablenotemark{c} & & 
0.51 (0.32--0.74) & 0.31 (0.27--0.37) \\ 
\hline
Abundance & & & \\
~~Ne & & 0.28 ($<$0.88) & 0 ($<$0.52) \\
~~Mg & & 0.94 (0.65--1.29) & 1.17 (1.04--1.58) \\
~~Si & & 0.50 (0.37--0.58) & 0.65 (0.55--0.77) \\
~~S & & 0.10 ($<$0.29) & 0.54 (0.30--0.83) \\
~~Fe & & 0.53 (0.48--0.58) & 0.62 (0.56--0.79) \\ 
\hline 
$\chi^2$/d.o.f. & & 137/129 & 162/180 \\ 
\enddata
\tablecomments{Each abundance is relative to the solar value and those of 
two plasma components were linked together. The 90\% confidence range
(c.r.) for temperature, ionization timescale, and abundances are given
in brackets.}
\tablenotetext{a}{The column density of an interstellar absorption assuming 
solar abundances.}
\tablenotetext{b}{Emission Measure: the unit is 
4$\pi d^2 \times 10^{12}$~cm$^{-5}$ where $d$ is the distance to W28.}
\tablenotetext{c}{This factor represents the ratio of the high-T emission 
measure to the low-T one.}
\end{deluxetable}
\begin{deluxetable}{lc}
\tablecaption{The best-fit parameters for the 3C~391 spectrum
\label{3c391param}}
\tablehead{
Parameter [unit] & Entire region (90\% c.r.) }
\startdata
\Nh\ [10$^{22}$~cm$^{-2}$] & 2.99 (2.89--3.10) \\
\kTe\ [keV] & 0.53 (0.50--0.57) \\
$n_{\mathrm{e}} t$ [10$^{12}$ cm$^{-3}$~s] & 2.5 ($>$ 0.9) \\
Mg & 1.24 (1.03--1.48) \\
Si & 1.12 (0.98--1.26) \\
S & 0.81 (0.56--1.07) \\
E.M.\tablenotemark{\P} & 14.4 (11.9--17.8) \\ \hline
$\chi^2$/d.o.f & 93/83 
\enddata
\tablecomments{Each abundance is relative to the solar value. 
The 90\% confidence range (c.r.) are given in brackets.}
\tablenotetext{\P}{Emission Measure: the unit is 
4$\pi d^2 \times 10^{12}$~cm$^{-5}$ where $d$ is the distance to 3C391.}
\end{deluxetable}

\begin{deluxetable}{lc}
\tablecaption{The best-fit parameters for the Kes~27 spectrum
\label{kes27param}}
\tablehead{
Parameter [unit] & Entire region (90\% c.r.) }
\startdata
\Nh\ [10$^{22}$~cm$^{-2}$] & 2.2 (2.0--2.3) \\
\kTe\ [keV] & 0.94 (0.84--1.01) \\
$n_{\mathrm{e}} t$ [10$^{12}$ cm$^{-3}$~s] & 0.7 ($>$ 0.3) \\
Mg & 1.0 (0.6--1.8) \\
Si & 1.4 (1.3--2.0) \\
S & 1.2 (0.8--1.6) \\
E.M.\tablenotemark{\P} & 3.09 (2.52--3.40) \\ \hline
$\chi^2$/d.o.f & 64/64 \\
\enddata
\tablecomments{Each abundance is relative to the solar value. 
The 90\% confidence range (c.r.) are given in brackets.}
\tablenotetext{\P}{Emission Measure: the unit is 
4$\pi d^2 \times 10^{12}$~cm$^{-5}$ where $d$ is the distance to
Kes~27.}
\end{deluxetable}

\begin{deluxetable}{lccc}
\tablecaption{Electron temperatures and densities of mixed-morphology SNRs 
and Cygnus Loop
\label{param_mmsnr}}
\tablehead{
SNR & region & electron temperature & density \\ & & [keV] & [cm$^{-3}$]
}
\startdata
\ic\ & interior & $1.08\pm0.03$ & 1.0 \\
 & exterior & $0.19\pm0.02$ & 4.2 \\ 
W49B & interior & 1.70$^{+0.02}_{-0.04}$ & 2.5 \\
 & exterior & 0.24$^{+0.04}_{-0.02}$ & 18 \\
W28 & center, southwest & $1.41^{+0.25}_{-0.10}$
& 0.25\tablenotemark{(1)} \\
 & center, northeast & $0.62\pm0.03$
& 0.75\tablenotemark{(1)} \\
W44 & center & $0.84^{+0.02}_{-0.05}$ 
& 0.4\tablenotemark{(2)}~~~\tablenotemark{\dag} \\ 
& north & $0.64^{+0.04}_{-0.05}$ & - \\
3C391 & entire & $0.53^{+0.04}_{-0.03}$ & 1.1 \\
Kes~27 & center & $0.84\pm0.08$\tablenotemark{(3)} 
& 0.16\tablenotemark{\dag} \\
 & rim & $0.59^{+0.04}_{-0.06}$\tablenotemark{(3)} & - \\
\hline
Cygnus Loop & center & $0.82^{+0.09}_{-0.07}$\tablenotemark{(4)} & - \\
 & northeast limb &  0.22--0.30\tablenotemark{(5)} & - \\
\enddata
\tablenotetext{\dag}{mean density of the entire remnant.}
\tablerefs{(1)\citet{rho02}; (2)\citet{harrus97};
(3)\citet{enoguchi02}; (4)\citet{miyata98}; (5)\citet{miyata94}.}
\end{deluxetable}

\begin{deluxetable}{lccccc}
\tablecaption{Electron and ionization temperatures of mixed-morphology SNRs
\label{ion_mmsnr}
}
\tablehead{
SNR name & (region) & & \kTe\ & \kTz\ & Element\tablenotemark{b} \\ 
& & & [keV] & [keV] & \\
}
\startdata
\ic\tablenotemark{a} & (center) & 
& $0.99^{+0.04}_{-0.05}$ & $1.49^{+0.09}_{-0.10}$ & S \\
W49B\tablenotemark{a} & (entire) & 
& $1.77^{+0.04}_{-0.03}$ & $2.47^{+0.21}_{-0.20}$ & Ca \\
W28\tablenotemark{a} & (east) &
& $1.07^{+0.43}_{-0.28}$ & $0.93^{+0.33}_{-0.31}$ & Si \\
W44 & (center) & & $0.78^{+0.03}_{-0.04}$ 
& $0.68^{+0.05}_{-0.04}$ & Si \\
3C391 & (entire) & & $0.53^{+0.03}_{-0.04}$ 
& $0.47^{+0.03}_{-0.05}$ & Mg \\
Kes~27 & (entire) & & $0.82^{+0.19}_{-0.16}$ 
& $0.83^{+0.11}_{-0.12}$ & Si \\ 
\enddata
\tablenotetext{a}{Derived temperatures are for the high-temperature plasma.}
\tablenotetext{b}{This column shows elements used to estimate the 
ionization temperature of the plasma.}
\end{deluxetable}

\begin{deluxetable}{lcccccc}
\tablecaption{Electron and ionization temperatures of shell-like SNRs 
\label{ion_shellsnr}}
\tablehead{
SNR name & & \kTe\ & $\log (n_\mathrm{e} t)$ & \kTz\ & Element & reference \\
 & & [keV] & [\ntunit] & [keV] & 
}
\startdata
Cassiopeia~A & & $3.02^{+0.30}_{-0.24}$ & $10.83^{+0.02}_{-0.04}$ 
& $1.07^{+0.02}_{-0.04}$ & S & our work \\
Kepler's SNR & & $2.38\pm0.14$ & $10.84\pm0.03$ & $0.97\pm0.03$ & S & 1 \\
Tycho's SNR\tablenotemark{a} 
& & $0.85\pm0.15$ & $11.0\pm0.1$ & $0.57\pm0.03$ & Si & 2 \\
G352.7$-$0.1 & & $2.0^{+1.6}_{-0.7}$ & $11.0^{+0.5}_{-0.3}$ 
& $1.00^{+0.46}_{-0.22}$ & S & 3 \\
Puppis~A & & $0.72\pm0.06$ & $10.47\pm0.07$ & 
$0.353^{+0.011}_{-0.008}$ & Mg & 4 \\
RCW~86 & & $0.430^{+0.015}_{-0.017}$ & $10.71\pm0.05$
& $0.286^{+0.005}_{-0.004}$ & Mg & 5 \\
Cygnus Loop\tablenotemark{b} 
& & $0.82^{+0.09}_{-0.07}$ & $10.35^{+0.06}_{-0.07}$
& $0.54^{+0.05}_{-0.06}$ & Si & 6 \\
\enddata
\tablerefs{(1)\citet{kinugasa99}; (2)\citet{hwang97}; (3)\citet{kinugasa98};
(4)\citet{tamuraphd}; (5)\citet{rho02b}; (6)\citet{miyata98}.}
\tablenotetext{a}{For Tycho's SNR, both \kTe\ and $n_\mathrm{e} t$ 
are derived from Si and S line diagnostics, not from NEI model
fittings. Details are described in \citet{hwang97}.}
\tablenotetext{b}{Listed \kTe\ and $n_\mathrm{e} t$ are those in the 
central region of the remnant, see \citet{miyata98}.}
\end{deluxetable}


\end{document}